\documentclass[12pt,draftcls,onecolumn]{IEEEtran}
\ifCLASSINFOpdf
\else
\fi
\usepackage{url}

\usepackage{bm,bbm,amssymb,mathrsfs,mathtools,enumitem,cite,setspace}

\usepackage{xcolor}

\newtheorem{theorem}{Theorem}

\newtheorem{definition}{Definition}

\newtheorem{lemma}{Lemma}
\newtheorem{corollary}{Corollary}
\newtheorem{proposition}{Proposition}

\newtheorem{observation}{Observation}
\newtheorem{assumption}{Assumption}

\newtheorem{remark}{Remark}
\newtheorem{criterion}{Criterion}

\newcommand{\R}{{\mathbb R}}

\newcommand{\N}{{\mathbb N}}
\newcommand{\E}{{\mathbb E}}
\newcommand{\X}{{\mathcal X}}
\newcommand{\U}{{\mathcal U}}

\newcommand{\Py}{{\mathbb P}}

\newcommand{\vct}[1]{{#1}}

\newcommand{\rv}[1]{{{#1}}}
\newcommand{\I}{{\mathscr{I}}}

\DeclareMathOperator*{\argmax}{arg\,max}

\begin{document}
%
\title{Fixed-horizon Active Hypothesis Testing}
%
%
%

\author{Dhruva~Kartik,
        Ashutosh~Nayyar,~\IEEEmembership{Senior Member,~IEEE,}
        and~Urbashi~Mitra,~\IEEEmembership{Fellow,~IEEE}
\thanks{Preliminary version of this paper titled ``Active hypothesis testing: beyond Chernoff-Stein'' was presented at the 2019 IEEE International Symposium on Information Theory.}
\thanks{Dhruva Kartik, Ashutosh Nayyar and Urbashi Mitra are with the Ming Hsieh Department
of Electrical and Computer Engineering, University of Southern California, Los Angeles,
CA, 90007 USA. E-mail: mokhasun, ashutosh.nayyar, ubli@usc.edu.}
\thanks{Manuscript received April 19, 2005; revised August 26, 2015.}}

%
%

\markboth{Submitted to IEEE Transactions on Automatic Control}%
{Shell \MakeLowercase{\textit{et al.}}: Bare Demo of IEEEtran.cls for IEEE Journals}
%



\maketitle

\begin{abstract}
Two active hypothesis testing problems are formulated. In these problems, the agent can perform a fixed number of experiments and then decide on one of the hypotheses. The agent is also allowed to declare its experiments inconclusive if needed. The first problem is an asymmetric formulation in which the the objective is to minimize the probability of incorrectly declaring a particular hypothesis to be true while ensuring that the probability of correctly declaring that hypothesis is moderately high. This formulation can be seen as a generalization of the formulation in the classical Chernoff-Stein lemma to an active setting. The second problem is a symmetric formulation in which the objective is to minimize the probability of making an incorrect inference (misclassification probability) while ensuring that the true hypothesis is declared conclusively with moderately high probability. For these problems, lower and upper bounds on the optimal misclassification probabilities are derived and these bounds are shown to be asymptotically tight. Classical approaches for experiment selection suggest use of randomized and, in some cases, open-loop strategies. As opposed to these classical approaches, fully deterministic and adaptive experiment selection strategies are provided. It is shown that these strategies are asymptotically optimal and further, using numerical experiments, it is demonstrated that these novel experiment selection strategies (coupled with appropriate inference strategies) have a significantly better performance in the non-asymptotic regime.
\end{abstract}


%
\IEEEpeerreviewmaketitle

\section{Introduction}
%
%
%
%

\IEEEPARstart{W}{e} frequently encounter scenarios wherein we would like to deduce whether one of several hypotheses is true by gathering data or evidence. This problem is referred to as multi-hypothesis testing. If we have access to multiple candidate experiments or data sources, we can adaptively select more informative experiments to infer the true hypothesis. This leads to a joint control and inference problem commonly referred to as \emph{active} hypothesis testing. There are numerous ways of formulating this problem and the precise mathematical formulation depends on the target application.

In this paper, we consider a scenario in which there is an agent that can perform a \emph{fixed} number of experiments. Subsequently, the agent can decide on one of the hypotheses using the collected data. The agent is also allowed to declare the experiments \emph{inconclusive} if needed. In this fixed-horizon setting, we consider two formulations. In the first formulation, we are interested in minimizing the probability of incorrectly declaring a \emph{particular} hypothesis to be true while ensuring that the probability of correctly declaring the same hypothesis is moderately high. Thus, we would like to declare that this hypothesis is true only if we have very strong evidence supporting it. This formulation is intended for applications like anomaly detection wherein incorrectly declaring the system to be safe (\emph{i.e.} anomaly-free) can be very expensive whereas a moderate number of false alarms can be tolerated. This formulation, and thus our results, can be viewed as a generalization of the classical Chernoff-Stein lemma \cite{cover2012elements} to an active multi-hypothesis testing setup.

In the second formulation, we are interested in minimizing the probability of making an
\emph{incorrect} inference (misclassification probability) while ensuring
that the true hypothesis is declared conclusively with moderately
high probability. The key difference between the first and second formulations is that the former is asymmetric, \emph{i.e.}, it focuses on reliably inferring a \emph{particular} hypothesis, whereas the latter formulation is symmetric in the sense that it aims to avoid misclassifying every hypothesis. This symmetric formulation is of particular interest when the penalty for making \emph{any} incorrect inference is significantly higher than the penalty for making no decision. In such cases, it is reasonable for the agent to abstain from drawing conclusions unless there is strong evidence supporting one of the hypotheses. 

In both these problems, the agent can select experiments at each time in a data-driven manner. We refer to the strategy used for selecting these experiments as the \emph{experiment selection strategy}. We refer to the strategy used by the agent to make an inference (or to declare its experiments inconclusive) based on all the data collected as the \emph{inference strategy}. Thus, the two problems described above involve optimization over the space of inference and experiment selection strategy pairs.


Our contributions in this paper pertaining to these hypothesis testing problems can be summarized as follows.
\begin{enumerate}
\item We find lower and upper bounds on the optimal misclassification probabilities in our constrained optimization problems. These bounds are asymptotically (w.r.t. the time-horizon) tight under some mild assumptions. Thus, we characterize the optimal misclassification error exponents in each problem.
\item We propose a novel approach for designing experiment selection strategies. Unlike the classical approach which results in randomized and, in some cases, open-loop strategies, this approach allows us to design deterministic and adaptive experiment selection strategies that are asymptotically optimal.
\item We demonstrate numerically that the experiment selection strategies designed using our approach, when coupled with appropriate inference strategies, achieve superior non-asymptotic performance in comparison to the classical approaches.
\end{enumerate}

The rest of the paper is organized as follows. In Section \ref{priorwork}, we summarize key prior literature on hypothesis testing and discuss how our problem is related to various other formulations. In Section \ref{notation}, we describe our notation. We describe our system model in Section \ref{system} and in Section \ref{probformsub}, we formulate our problems. We state the main results in Section \ref{mainresults} and sketch the proof of our results. We provide a detailed analysis of our hypothesis testing problems in Sections \ref{sec:asymmetric} and \ref{thm3proof}. In Section \ref{sec:example}, we discuss an anomaly detection example and provide the results of our numerical experiments. We conclude the paper in Section \ref{conc}.

\subsection{Related Work}\label{priorwork}
Hypothesis testing is a long-standing problem and has been addressed in various settings. Works that are closely related to active hypothesis testing can be broadly classified into the following paradigms.
\subsubsection{Fixed-horizon Hypothesis Testing}
In the simplest fixed-horizon hypothesis testing setup, we have binary hypotheses and a single experiment. The inference is made based on a fixed number of i.i.d. observations obtained by repeatedly performing this experiment. In this setup, there are two popular formulations: (i) the Neyman-Pearson type asymmetric formulation used in the Chernoff-Stein lemma \cite{cover2012elements}; (ii) the unconstrained symmetric formulation that involves minimizing the Bayesian error probability \cite{cover2012elements}. While our asymmetric formulation is a generalization of the Neyman-Pearson type formulation, our symmetric formulation is different from Bayesian error probability minimization in \cite{cover2012elements}. The key difference is that in Bayesian error minimization, the agent is not allowed to declare its experiments inconclusive at the end of the horizon and \emph{must} declare one of the hypotheses to be true. More general works in this paradigm include \cite{blahut1974hypothesis, blahut1976information,tuncel2005error,  liu2018second,polyanskiy2010channel}. All the aforementioned works are passive in the sense that there is only one experiment and thus, the experiment selection strategy is trivial. Nevertheless, we employ many of the analysis techniques developed in these works all of which are available in the form of lecture notes in \cite{polyanotes}.

An active fixed-horizon formulation has been considered in \cite{nitinawarat2013controlled} in which the objective is to minimize the maximal error probability. This formulation is symmetric and does not allow the inconclusive declaration. Allowing the inconclusive declaration makes the nature of our analysis and results significantly different from the formulation in \cite{nitinawarat2013controlled}.

\subsubsection{Sequential Hypothesis Testing} In sequential hypothesis testing, the time horizon is not fixed and the agent can continue to perform experiments until a stopping criterion is met. The objective then is to minimize a linear combination of the expected stopping time and the Bayesian error probability.
Inspired by Wald's sequential probability ratio test (SPRT) \cite{wald1973sequential}, Chernoff first addressed the problem of \emph{active} sequential hypothesis testing in \cite{chernoff1959sequential}. This work was later generalized in \cite{bessler1960theory,nitinawarat2013controlled,naghshvar2013active}. Although our formulations have a fixed time-horizon, they are most closely related to the sequential active hypothesis testing framework. Intuitively, this is because in both the sequential setting and our fixed-horizon setting, the agent conclusively declares a hypothesis to be true only if there is strong evidence supporting it. If strong enough evidence is not found, the agent in the sequential setting continues to perform experiments whereas the agent in our setting simply declares the experiments inconclusive. Fixed-horizon formulations are useful in applications with hard time constraints and the agent does not have the luxury to keep performing experiments until strong enough evidence is obtained.

Because of the strong parallels between our setting and the sequential setting, we will use the strategies developed in the sequential setting as benchmarks in our numerical experiments. In all the aforementioned works on the sequential setting, the experiment selection strategy has a randomized component. Although these randomized strategies are asymptotically optimal, their non-asymptotic performance may be poor. Deterministic strategies were proposed in \cite{chernoff1959sequential, naghshvar2012extrinsic} but in many cases, these strategies are \emph{not} asymptotically optimal. In this paper, we develop an approach that helps us in designing deterministic, adaptive and asymptotically optimal experiment selection strategies. Moreover, in some scenrios like anomaly detection, our deterministic strategies have a significantly better non-asymptotic performance. {The proof of asymptotic optimality of our deterministic strategies is for the fixed-horizon setting. However, in some special cases, it may be possible to extend the proof to the sequential setting by means of Lemmas 2 and 3 in \cite{chernoff1959sequential}.}

\subsubsection{Anomaly Detection}
Many anomaly detection problems can be viewed as active hypothesis testing problems. In anomaly detection, there are multiple \emph{normal} processes which exhibit a certain kind of statistical behavior. Among these processes, there could be an anomaly with statistical characteristics distinct from the normal processes. There are various mechanisms to probe these processes and the objective is to reliably detect the anomaly as quickly as possible. Some recent works in anomaly detection include \cite{huang2018active,chen2019active,tsope,gure,chao}. All these works are in the sequential setting. It has been noted in \cite{huang2018active, tsope, chen2019active} that deterministic strategies achieve better performance. We believe that these deterministic strategies may be related to ours.

The problem of oddball detection has been considered in \cite{oddball,neuraloddball}. The approach used in \cite{oddball,neuraloddball} is similar to Chernoff's approach in \cite{chernoff1959sequential} but the key innovation is that they do not assume knowledge of the underlying distributions.

\subsubsection{Variable-length Communication with Feedback}
This problem is concerned with designing variable-length codes for discrete memoryless communication channels with perfect feedback \cite{burnashev1976data,naghshvarvariable}. From the receiver's perspective, this problem can be viewed as a sequential hypothesis testing problem \cite{naghshvarvariable}. The receiver aims to decode the message (unknown hypothesis) based on the received symbols (observations). At each time, based on all the past received symbols, the receiver selects a \emph{mapping} that maps the message to the channel input. The transmitter then simply uses this \emph{mapping} to encode the message at the next time step. These mappings are the receiver's experiments. 
Our framework allows us to formulate a fixed-horizon analogue of the variable-length coding problem. In this fixed-horizon formulation, the receiver is allowed to declare the transmission inconclusive if needed. The objective of the receiver is to minimize the probability of incorrectly decoding the message while satisfying a constraint on the probability of correctly decoding the message.

\subsection{Notation}\label{notation}
Random variables are denoted by upper case letters ($X$), their realization by the corresponding lower case letter ($x$). We use calligraphic fonts to denote sets ($\mathcal{U}$). The probability simplex over a finite set $\mathcal{U}$ is denoted by $\Delta \mathcal{U}$. In general, subscripts denote time indices unless stated otherwise. For time indices $n_1\leq n_2$, $\rv{Y}_{n_1:n_2}$ denotes the collection of variables $(\rv{Y}_{n_1},\rv{Y}_{n_1+1},...,\rv{Y}_{n_2})$.
For a strategy $g$, we use $\Py^g[\cdot]$ and $\E^g[\cdot]$ to indicate that the probability and expectation depend on the choice of $g$. For an hypothesis $i$, $\E_i^g[\cdot]$ denotes the expectation conditioned on hypothesis $i$. {For a random variable $X$ and an event $\mathcal{E}$, $\E[X ; \mathcal{E}]$ denotes $\E[X\mathbbm{1}_{\mathcal{E}}]$, where $\mathbbm{1}_{\mathcal{E}}$ is the indicator function associated with the even $\mathcal{E}$.} The \emph{cross-entropy} between two distributions $p$ and $q$ over a finite space $\mathcal{Y}$ is given by
\begin{align}
H(p,q) = -\sum_{y \in \mathcal{Y}}p(y)\log q(y).
\end{align}
The \emph{Kullback-Leibler} divergence between distributions $p$ and $q$ is given by
\begin{equation}
D(p || q) = \sum_{y \in \mathcal{Y}}p(y)\log\frac{p(y)}{q(y)}.
\end{equation}

\section{Minimum Misclassification Error Problems}\label{sec:mis}
In this section, we will formulate the two active hypothesis testing problems. We will describe our assumptions and state our main results on the asymptotic behavior of optimal misclassification probabilities.
\subsection{System Model}\label{system}
Let $\mathcal{X} = \{0,1,\ldots,M-1\}$ be a finite set of hypotheses and let the random variable $\rv{X}$ denote the true hypothesis. The prior probability on $\rv{X}$ is $\vct{\rho}_1$. Without loss of generality, let us assume that the distribution $\rho_1$ has full support. At each time $n=1,2,\ldots$, an agent can perform an experiment $\rv{U }_n \in \mathcal{U}$ and obtain an observation $\rv{Y}_n \in \mathcal{Y}$. {We assume that the sets $\mathcal{U}$ and $\mathcal{Y}$ are finite.}  The observation $\rv{Y}_n$ at time $n$ is given by
\begin{equation}\label{obs}
\rv{Y}_n = \xi(\rv{X}, \rv{U}_n,\rv{W}_n),
\end{equation}
where $\{\rv{W}_n:  n = 1,2,\dots\}$ is a collection of mutually independent and identically distributed \emph{primitive} random variables. The probability of observing  $y$ after performing an experiment $u$ under hypothesis $i$ is denoted by $p_i^u(y)$,
\[ p_i^u(y) \doteq \Py(\rv{Y}_n=y \mid \rv{X}=i,\rv{U}_n=u).  \]
 The time horizon, that is, the total number of experiments performed is fixed \emph{a priori} to $N < \infty$.

At time $n=1,2,\ldots, N$, the information available to the agent, denoted by $\rv{I}_n$, is the collection of all experiments performed and the corresponding observations up to time $n-1$, 
\begin{equation}
\rv{I}_n \doteq \{\rv{U}_{1:n-1},\rv{Y}_{1:n-1}\}.
\end{equation}
Let the collection of all possible realizations of information $\rv{I}_n$ be denoted by $\mathcal{I}_n$. At time $n$, the agent selects a distribution over the set of actions $\mathcal{U}$ according to an \emph{experiment selection rule} $g_n: \mathcal{I}_n \to \Delta \mathcal{U}$ and the action $\rv{U}_n$ is randomly drawn from the distribution $g_n(\rv{I}_n)$,
that is,
\begin{equation}
\rv{U}_n  \sim g_n(\rv{I}_n).
\end{equation}
{For a given experiment $u \in \mathcal{U}$ and information realization $\mathscr{I} \in \mathcal{I}_n$, the probability $\Py^g[U_n = u \mid I_n = \mathscr{I}]$ is denoted by $g_n(\mathscr{I} : u)$.}
The sequence  $\{g_n, n=1,\ldots,N\}$ is denoted by $g$ and  referred to as the \emph{experiment selection strategy}. Let the collection of all such experiment selection strategies be $\mathcal{G}$. 

After performing $N$ experiments, the agent can declare one of the hypotheses  to be true or it can declare that its experiments were inconclusive. We refer to this final declaration as the  agent's \emph{inference decision} and denote it by  $\hat{\rv{X}}_N$.  Thus, the inference decision can take values in $\mathcal{X} \cup \{\aleph\}$, where $\aleph$ denotes the inconclusive declaration. Using the information $I_{N+1}$, the agent chooses a distribution over the set of hypotheses according to an \emph{inference strategy} $f: \mathcal{I}_{N+1} \to \Delta(\mathcal{X} \cup \{\aleph\})$ and the inference $\hat{\rv{X}}_N$ is drawn from the distribution $f(\rv{I}_{N+1})$,  \emph{i.e}.
\begin{equation}
    \hat{\rv{X}}_{N} \sim f(\rv{I}_{N+1}).
\end{equation}
{For a given inference $\hat{x} \in \mathcal{X} \cup \{\aleph\}$ and information realization $\mathscr{I} \in \mathcal{I}_{N+1}$, the probability $\Py^{f,g}[\hat{X}_N = u \mid I_{N+1} = \mathscr{I}]$ is denoted by $f_N(\mathscr{I} : \hat{x})$.} Let the set of all inference strategies be $\mathcal{F}$. 
For an experiment selection strategy $g$ and an inference strategy $f$, we define the following probabilities.
\begin{definition}
For $i \in \mathcal{X}$, let $\psi_N(i)$ be the probability that the agent infers hypothesis $i$ given that the true hypothesis is $i$, \emph{i.e}.
\begin{align}
    \psi_N(i) &\doteq \Py^{f,g}[\hat{\rv{X}}_{N} = i \mid \rv{X} = i].\\
    \shortintertext{We refer to $\psi_N(i)$ as the \emph{correct-inference probability} of type-$i$. Let $\phi_N(i)$ be the probability that the agent infers $i$ given that the true hypothesis is not $i$, \emph{i.e}.}
    \phi_N(i) &\doteq \Py^{f,g}[\hat{\rv{X}}_{N} = i \mid \rv{X} \neq i].
\end{align}
We refer to $\phi_N(i)$ as the \emph{misclassification probability} of type-$i$.
\end{definition}

We will also be interested in the event that the agent declares an incorrect hypothesis to be true.  That is, we will consider the event $\cup_{i \in \mathcal{X}} \{\hat{\rv{X}}_{N} = i, \rv{X} \neq i \}$. We refer to this event as the \emph{misclassification event}.

\begin{definition}
Let $\gamma_N$ be the probability of making an incorrect inference, \emph{i.e.}
\begin{align}
    \gamma_N & \doteq \Py^{f,g}[\cup_{i \in \mathcal{X}} \{\hat{\rv{X}}_{N} = i, \rv{X} \neq i \}].
\end{align}
We will refer to $\gamma_N$ as the \emph{misclassification probability}.
\end{definition}
\begin{remark}
Note that the misclassification probability $\gamma_N$ can be expressed in terms of the misclassification probabilities $\phi_N(i)$ of type-$i$ in the following manner
\begin{align}
    \gamma_N &=  \sum_{i \in \mathcal{X}}\Py^{f,g}[\hat{\rv{X}}_{N} = i \mid \rv{X} \neq i]\Py[\rv{X} \neq i]\\
    &= \sum_{i \in \mathcal{X}}\phi_N(i)(1-\rho_1(i)).
\end{align}
\end{remark}


\subsection{Problem Formulation and Preliminaries}\label{probformsub}
We will consider two active hypothesis testing formulations. The first one is an asymmetric formulation in which the focus is on a particular hypothesis $i$ and involves minimizing the misclassification probability $\phi_N(i)$ of type-$i$. The second formulation is a symmetric formulation that involves minimizing the misclassification probability $\gamma_N$.

\subsubsection{The Asymmetric Formulation (\ref{opti})}
In this formulation, we are interested in designing an experiment selection strategy $g$ and an inference strategy $f$ that minimize the misclassification probability $\phi_N(i)$ of type-$i$ subject to the constraint that the correct-inference probability $\psi_N(i)$ of type-$i$ is sufficiently large.
In other words, we would like to solve the following optimization problem: 
\begin{align}\tag{P1}\label{opti}
    & & & \underset{f \in \mathcal{F},g \in \mathcal{G}}{\text{inf}} & & \phi_N(i) & \\
    \nonumber & & & \text{subject to} & & \psi_N(i) \geq 1-\epsilon_N, & 
\end{align}
where $0 < \epsilon_N < 1$. Let the infimum value of this optimization problem be $\phi^*_N(i)$. 
Note that this problem is always feasible because the agent can trivially satisfy the correct-inference probability constraint by always declaring hypothesis $i$. We refer to this problem as the \emph{ minimum misclassification error problem of type-$i$} or simply Problem (\ref{opti}).
\begin{remark}
Problem (\ref{opti}) can be seen as a binary hypothesis testing problem with null hypothesis $\{X = i\}$ and alternate hypothesis $\{X \neq i\}$. We observe that when there is only one experiment and two hypotheses, this formulation is identical to that of the Chernoff-Stein lemma in \cite{cover2012elements}.
\end{remark}

This formulation is helpful in modeling scenarios in which the cost of incorrectly declaring a particular hypothesis to be true is very high. For instance, consider a system which can potentially have various types of anomalies. We are interested in testing whether the system has no anomalies (hypothesis $X = i$) or has some anomaly (hypothesis $X \neq i$). In such systems, a few false alarms may be tolerable but declaring that the system is free of anomalies when there is one can be very expensive. Therefore, we would like to minimize the probability of falsely declaring the system to be free of anomalies subject to the constraint that the probability of raising false alarms is sufficiently small. Clearly, this scenario can be modeled using the asymmetric formulation (\ref{opti}).

\subsubsection{The Symmetric Formulation (\ref{opt1})}
In this formulation, we are interested in designing an experiment selection strategy $g$ and an inference strategy $f$ that minimize the misclassification probability $\gamma_N$ while satisfying the constraint that the correct-inference probability $\psi_N(i)$ of type-$i$ is sufficiently large for every hypothesis $i \in \mathcal{X}$. In other words, we would like to solve the following optimization problem:
\begin{align}\tag{P2}\label{opt1}
    & & & \underset{f \in \mathcal{F},g \in \mathcal{G}}{\text{min}} & & \gamma_N & \\
    \nonumber& & & \text{subject to} & & \psi_N(i) \geq 1-\epsilon_N,\; \forall i \in \mathcal{X}, & 
\end{align}
where $0 < \epsilon_N < 1$. Let $\gamma^*_N$ denote the infimum value of this optimization problem. We define $\gamma^*_N \doteq \infty$ if the optimization problem is infeasible. We refer to this problem as the \emph{minimum misclassification error problem} or simply Problem (\ref{opt1}).

The above  formulation is intended for scenarios where the penalty for declaring an incorrect hypothesis to be true is much higher than the penalty for making no decision about the hypothesis. In such cases, it is reasonable for the agent to abstain from drawing conclusions when the evidence is not strong enough. The constraints on type-$i$ correct-inference probabilities $\psi_N(i)$ ensure that the agent does not abstain from drawing conclusions too often. Thus, the optimization problem (\ref{opt1}) aims to find experiment selection and inference strategies that misclassify the least among all those strategies that make the correct inference with high probability.

%

\begin{definition}[Log-likelihood ratio]
For an experiment $u \in \mathcal{U}$ and any pair of hypotheses $i,j \in \mathcal{X}$ let
\begin{align}\label{llrdef}
\lambda_j^i(u,y) \doteq \log\frac{p_i^u(y)}{p_j^u(y)}
\end{align}
be the log-likelihood ratio associated with an observation $y \in \mathcal{Y}$.
\end{definition}

We make the following assumptions on our system model.
\begin{assumption}[\emph{Common Support}]\label{boundedassump}
For any given experiment $u \in \mathcal{U}$, there exists a non-empty set $\mathcal{Y}(u) \subseteq \mathcal{Y}$ such that for every hypothesis $i \in \mathcal{X}$, the support of the distribution $p_i^u$ is $\mathcal{Y}(u)$. In other words, for every hypothesis $i \in \mathcal{X}$, $p_i^u(y) > 0$ if and only if $y \in \mathcal{Y}(u)$.
\end{assumption}

Let $B >0$ be a constant such that $|\lambda_j^i(u,y)| < B$ for every experiment $u \in \mathcal{U}$, observation $y \in \mathcal{Y}(u)$ and any pair of hypotheses $i,j \in \mathcal{X}$. Note that the existence of such a constant $B$ is guaranteed because of Assumption \ref{boundedassump}.

Recall that the observation space $\mathcal{Y}$ is finite in our model. In that case, our Assumption \ref{boundedassump} is equivalent to the assumption that the variance $\E_i[(\lambda^i_j(u,Y))^2] - (\E_i[\lambda^i_j(u,Y)])^2$ of the log-likelihood ratio $\lambda^i_j$ is finite for every experiment $u$ and pair of hypotheses $i,j$. This finite variance assumption is standard in the active hypothesis testing literature \cite{chernoff1959sequential,nitinawarat2013controlled,naghshvar2013active}. 
\begin{remark}
In this paper, we restrict our analysis to the setting where the observation space $\mathcal{Y}$ is finite. However, we believe that our analysis can be extended to the setting of \cite{nitinawarat2013controlled} which allows infinite observation spaces with an additional assumption that the log-likelihood ratios are sub-Gaussian.
\end{remark}

\begin{assumption}\label{epsassum}
We have that the bound $1-\epsilon_N$ on the type-$i$ correct-inference probability {satisfies $\epsilon_N \to 0$}. Further,
\begin{align}
    \lim_{N \to \infty}\frac{-\log{\epsilon_N}}{N} = 0.
\end{align}
\end{assumption}
This assumption simply captures the fact that we need the hit probabilities to be large, but not necessarily too large.
The following assumption is made for ease of exposition and is intended only for the symmetric formulation (\ref{opt1}). It is not required for the asymmetric formulation (\ref{opti}).
\begin{assumption}\label{steadinf}
For each experiment $u \in \mathcal{U}$ and any pair of hypotheses $i,j \in \mathcal{X}$ such that $i \neq j$, we have
\begin{align}
D(p_i^u || p_j^u) > 0.
\end{align}
\end{assumption}
We refer the reader to \cite{chernoff1959sequential,nitinawarat2013controlled,naghshvar2013active} for techniques that can be used to relax Assumption \ref{steadinf}.
Before stating our main results, we will define some important quantities.

\begin{definition}[Max-min Kullback-Leibler Divergence]
\label{kldef}
For each hypothesis $i \in \mathcal{X}$, define
\begin{align}
\label{maxmin}D^*(i) &\doteq \max_{\vct{\alpha} \in \Delta\mathcal{U}} \min_{j\neq i} \sum_{u}\alpha(u) D(p_i^u || p_j^u) \\
\label{minmax}&=  \min_{\vct{\beta} \in \Delta\tilde{\mathcal{X}}_i} \max_{u \in \mathcal{U}} \sum_{j \neq i}\beta(j) D(p_i^u || p_j^u),
\end{align}
where $\tilde{\mathcal{X}}_i = \mathcal{X} \setminus \{i\}$. Note that $\alpha$ is a distribution over the set of experiments $\mathcal{U}$ and $\beta$ is a distribution over the set of alternate hypotheses $\tilde{\mathcal{X}}_i$. The max-min Kullback-Leibler divergence $D^*(i)$ can be viewed as the value of a two-player zero-sum game \cite{chernoff1959sequential,osborne1994course}. In this zero-sum game, the maximizing player selects a mixed strategy $\alpha \in \Delta\U$ and the minimizing player selects a mixed strategy $\beta \in \Delta\tilde{\mathcal{X}}_i$. The payoff associated with these strategies is
\begin{align}
 \sum_{u}\sum_{j \neq i} \alpha(u)D(p_i^u || p_j^u)\beta(j).\label{zerosum}
\end{align}
The equality of the min-max and max-min values follows from the minimax theorem \cite{osborne1994course} because the sets $\mathcal{U}$ and $\mathcal{X}$ are finite and the Kullback-Leibler divergences are bounded by $B$ due to Assumption \ref{boundedassump}. Let the max-minimizer in (\ref{maxmin}) be denoted by $\alpha^{i*}$ and the min-maximizer in (\ref{minmax}) be denoted by $\beta^{i*}$.
\end{definition}

\begin{definition}[Posterior Belief]The \emph{posterior belief}  $\vct{\rho}_n$  on the hypothesis $\rv{X}$   based on information $\rv{I}_n$ is given by
\begin{equation}\label{postbelief}
\rho_n(i) = \Py[\rv{X} = i \mid \rv{U}_{1:n-1},\rv{Y}_{1:n-1}] =\Py[\rv{X} = i \mid \rv{I}_n].
\end{equation}
{Note that given a realization of the experiments and observations until time $n$,  the posterior belief does not depend on the experiment selection strategy $g$ or the inference strategy $f$.}
\end{definition}

\begin{definition}[Confidence Level]\label{confidencelevel}
For a hypothesis $i \in \mathcal{X}$ and a distribution $\rho$ on $\mathcal{X}$ such that $0 < \rho(i) < 1$, the \emph{confidence level} $\mathcal{C}_i(\vct{\rho})$ associated with hypothesis $i$ is defined as
\begin{equation}
\mathcal{C}_i(\vct{\rho}) \doteq \log\frac{\rho(i)}{1-\rho(i)}.
\end{equation}
The confidence level is the logarithm of the ratio of the probability (w.r.t. the distribution $\rho$) that hypothesis $i$ is true versus the probability that hypothesis $i$ is not true.
\end{definition}

\subsection{Main Results}\label{mainresults}

We will now state our \emph{three} main results on some asymptotic aspects of Problems (\ref{opti}) and (\ref{opt1}).

\subsubsection{Asymptotic Decay-Rate of Optimal Misclassification Probability $\phi_N^*(i)$ in Problem (\ref{opti})}\label{thm1subsec}
The following theorem can be viewed as a generalization of the classical Chernoff-Stein lemma \cite{cover2012elements} to the setting of active hypothesis testing. It states that the optimal value $\phi_N^*(i)$ in Problem (\ref{opti}) decays exponentially with the horizon $N$ and its asymptotic rate of decay is equal to the max-min Kullback-Leibler divergence $D^*(i)$ defined in Definition \ref{kldef}.
\begin{theorem}\label{p2asym}
The asymptotic rate of the optimal misclassification probability in Problem (\ref{opti}) is given by
\begin{align}
    \lim_{N \to \infty}-\frac{1}{N}\log\phi^*_N(i) =  D^*(i).
\end{align}
\end{theorem}

\emph{Proof sketch: }
To prove this result, we use the following approach. 
We first establish a lower bound on the misclassification probability $\phi_N(i)$ for any pair of experiment selection and inference strategies $(g,f)$ that satisfy the constraint $\psi_N(i) \geq 1-\epsilon_N$. The asymptotic decay-rate of this lower bound is equal to the max-min Kullback-Leibler divergence $D^*(i)$. We then construct experiment selection and inference strategies that asymptotically achieve this rate. The details of the proof of this result are included in Section \ref{sec:asymmetric}.
\hfill $\blacksquare$

The inference strategy constructed in the achievability proof of Theorem \ref{p2asym} has the following threshold structure
\begin{align}
\label{fNdef}{f}^N(\vct{\rho}_{N+1}:i) = 
\begin{cases}
1 &\text{if } \mathcal{C}_i(\vct{\rho}_{N+1}) - \mathcal{C}_i(\vct{\rho}_{1}) \geq  \theta_N\\
0 &\text{otherwise},
\end{cases}
\end{align}
where $\theta_N = ND^*(i) - o(N)$. The precise value of the threshold $\theta_N$ is provided in Appendix \ref{detthmproof}. The experiment selection strategy used is as follows: at each time $n$, randomly select an experiment $u$ from the max-min distribution $\alpha^{i*}$ as defined in Definition \ref{kldef}. This strategy design is motivated from the design in \cite{chernoff1959sequential}. Note that this is a completely open-loop strategy, that is, the experiments are selected without using any of the information acquired in the past. 

The open-loop randomized experiment selection strategy described above is asymptotically optimal for Problem (\ref{opti}). However, in the non-asymptotic regime, there may be other experiment selection strategies that perform significantly better than this open-loop randomized strategy. For some specialized problems such as anomaly detection, it was observed in recent active hypothesis testing literature \cite{huang2018active,tsope, chen2019active} that there exist deterministic and adaptive strategies that are asymptotically optimal and also, outperform the classical Chernoff-type randomized strategies in the non-asymptotic regime. Even in \cite{chernoff1959sequential}, Chernoff proposed a deterministic and adaptive strategy (see Section 7,  \cite{chernoff1959sequential}) but he also presented a counter-example in which the strategy was \emph{not} asymptotically optimal.

In this paper, we propose a class of deterministic and adaptive strategies that are asymptotically optimal for Problem (\ref{opti}). To the best of our knowledge, this is the first proof of asymptotic optimality of such strategies in a general active hypothesis testing setting. For a simple anomaly detection example, we also demonstrate numerically in Section \ref{sec:example} that this strategy (when paired with an appropriate inference strategy) performs significantly better than the open-loop randomized strategy $\alpha^{i*}$ in the non-asymptotic regime.

\subsubsection{{Asymptotically Optimal Experiment Selection Strategies for Problem (\ref{opti})}}\label{thm2subsec}

Let the moment generating function of the negative log-likelihood ratios $-\lambda_j^i(u,Y)$ for an experiment $u$ be denoted by $\mu_j^i(u,s)$, \emph{i.e.}
\begin{align}
\mu_j^i(u,s) \doteq \E_i[\exp\left(-s\lambda_j^i(u,Y)\right)].
\end{align}
\begin{definition} For a given experiment $u \in \mathcal{U}$, belief $\rho \in \Delta \mathcal{X}$ and $0 \leq s \leq 1$, define
\begin{align}
\mathscr{M}_i(u,\rho,s) &\doteq \frac{\sum_{j\neq i} ({{\rho}(j)})^s\mu_j^i(u,s)}{\sum_{j\neq i}  ({{\rho}(j)})^s}.
\end{align}
\end{definition}
Using this definition, we will now define a class of experiment selection strategies.

\begin{criterion}\label{crit} 
For a given time horizon $N$, consider an experiment selection strategy $g^N \doteq (g_1^N,g_2^N,\dots,g_N^N)$ such that at each time $n$, $\alpha_{n} \doteq g_n^N(I_n) \in \Delta \mathcal{U}$ satisfies
\begin{align}
\sum_{u \in \mathcal{U}} \alpha_n(u)\mathscr{M}_i(u,\rho_n,s_N) \leq \sum_{u \in \mathcal{U}} \alpha^{i*}(u)\mathscr{M}_i(u,\rho_n,s_N),
\end{align}
with probability 1. Here, $\rho_n$ is the posterior belief at time $n$ and 
\begin{align}
s_N &\doteq \min\left\{1,\sqrt{\frac{2\log\frac{M}{\epsilon_N}}{NB^2}}\right\}.\label{sdef}
\end{align}
\end{criterion}

\begin{observation} Criterion \ref{crit} captures three experiment selection strategies of interest. These are:
\begin{enumerate}
\item \emph{Open-loop Randomized Strategy (ORS):} At any time $n$, $g_n^N(I_n) \doteq \alpha^{i*}$. This is the open-loop randomized strategy discussed earlier in Section \ref{thm1subsec}.
\item \emph{Deterministic Adaptive Strategy (DAS):} At each time $n$, $g_n^N(I_n)$ selects the experiment $u \in \mathcal{U}$ that minimizes $\mathscr{M}_i(u,\rho_n,s_N)$. This is a deterministic and adaptive strategy.
\item \emph{Deterministic Adaptive Strategy with Restricted Support (DAS-RS):} At each time $n$, $g_n^N(I_n)$ selects the experiment $u$ from the \emph{support} of $\alpha^{i*}$ that minimizes $\mathscr{M}_i(u,\rho_n,s_N)$. This is also a deterministic and adaptive strategy.
\end{enumerate}
For all these experiment selection strategies, we have the following result.
\end{observation}

\begin{theorem}\label{detthm}
Let $f^N$ be as defined in \eqref{fNdef} and $g^N$ be an experiment selection strategy that satisfies Criterion \ref{crit}. Then the class of strategies $\{(f^N,g^N): N \in \N \}$ is asymptotically optimal. In other words, if $\psi_N(i)$ and $\phi_N(i)$ are the correct-inference and misclassification probabilities associated with the strategy pair $(f^N, g^N)$, then $\psi_N(i) \geq 1- \epsilon_N$ for every $N$ and
\begin{align}
\lim_{N \to \infty}-\frac{1}{N}\log\phi_N(i) = D^*(i).
\end{align} 
\end{theorem}
\begin{IEEEproof}
See Section \ref{achievable}.
\end{IEEEproof}

\begin{remark}[Zero-sum Game Interpretation]
Note that selecting an experiment that minimizes $\mathscr{M}_i(u,\rho,s)$ over the set $\mathcal{U}$ is equivalent to selecting an experiment that maximizes $(1-\mathscr{M}_i(u,\rho,s))/s$. When $s$ is small, this function can be approximated as follows
\begin{align}
\frac{1-\mathscr{M}_i(u,\rho,s)}{s} &= \frac{\sum_{j\neq i} ({{\rho}(j)})^s(1-\mu_j^i(u,s))}{s\sum_{j\neq i}  ({{\rho}(j)})^s}\\
&\approx \frac{\sum_{j\neq i} ({{\rho}(j)})^sD(p_i^u||p_j^u)}{\sum_{j\neq i}  ({{\rho}(j)})^s}\label{kldstrat},
\end{align}
since $(1-\mu_j^i(u,s))/s \to D(p_i^u||p_j^u)$ as $s \to 0$. Thus, we can interpret the strategy DAS in terms of the zero-sum game discussed earlier in the beginning of Section \ref{probformsub} after Definition \ref{kldef}. In the zero-sum game, if the minimizing player selects an alternate hypothesis $j$ with probability $\beta(j) = (\rho(j))^s/\sum_{k\neq i}(\rho(k))^s$, then the strategy DAS selects an approximate best-response to the minimizing player's strategy with respect to the payoff function in (\ref{zerosum}).
\end{remark}

Given a horizon $N$, the strategies DAS and DAS-RS described in this section are time-invariant. However, they depend on the value of $s_N$ and thus, on the horizon $N$ of the problem. In some cases, these strategies turn out to be independent of the value $s_N$ which results in fully stationary (with respect to the posterior belief $
\rho_n$) strategies. We show that this is indeed the case in the example discussed in Section \ref{sec:example}. It may be possible to show that for such stationary strategies, Lemma 2 in \cite{chernoff1959sequential} holds and thus, they are asymptotically optimal even in the \emph{sequential} formulation in \cite{chernoff1959sequential}.

\subsubsection{Asymptotic Decay-Rate of Optimal Misclassification Probability $\gamma_N^*$ in Problem (\ref{opt1})}\label{thm3subsec}
Similar to the result in Theorem \ref{p2asym}, we can characterize the decay-rate of the optimal misclassification probability $\gamma_N^*$ as follows.
\begin{theorem}\label{mainthm3}
The optimal misclassification probability $\gamma_N^*$ in Problem (\ref{opt1}) decays exponentially with the horizon $N$ and its asymptotic decay-rate is equal to $\min_{i \in \X}D^*(i)$, \emph{i.e.}
\begin{align}
\lim_{N \to \infty}-\frac{1}{N}\log\gamma_N^* = \min_{i\in \mathcal{X}}D^*(i).
\end{align}
\end{theorem}

\emph{Proof sketch: }
The methodology used for proving this result is very similar to that of Theorem \ref{p2asym}. We first obtain a lower bound on the misclassification probability $\gamma_N$ for any pair of experiment selection and inference strategies $(g,f)$ that satisfy the constraints of Problem \eqref{opt1}. This lower bound is very closely related to the lower bounds established for Problem (\ref{opti}). Then we construct a class of experiment selection and inference strategies that achieve this lower bound asymptotically. This class of experiment selection strategies includes the randomized strategy proposed in \cite{chernoff1959sequential} and also, deterministic strategies similar to DAS and DAS-RS introduced in Section \ref{thm2subsec}. The derivation of the lower bound and the construction of the experiment selection and inference strategies are discussed in detail in Section \ref{thm3proof}.
\hfill $\blacksquare$


\section{Analysis of Problem (\ref{opti})}\label{sec:asymmetric}

In this section, we analyze the asymmetric formulation (\ref{opti}). To optimize the misclassification error of type-$i$ in Problem (\ref{opti}), we need to design both an experiment selection strategy $g$ and an inference strategy $f$. We will first arbitrarily fix the experiment selection strategy $g$. For a fixed $g$, we derive lower bounds on the misclassification probability $\phi_N(i)$ associated with any inference strategy $f$ that satisfies the constraint in Problem (\ref{opti}). These bounds are obtained using the weak converse approach described in \cite{polyanotes}. In these derivations, we will introduce some useful properties of the confidence level defined in Definition \ref{confidencelevel}. Using these, we will then weaken these lower bounds to derive a bound that does not depend on the strategy $g$. Further, we show that any experiment selection strategy satisfying Criterion \ref{crit} defined in Section \ref{thm2subsec}, coupled with an appropriate inference strategy, can asymptotically achieve this strategy-independent lower bound. {Finally, we will discuss methods for obtaining better non-asymptotic lower bounds on the misclassification probability $\phi_N(i)$ using the strong converse theorem in \cite{polyanotes}.}

%

\subsection{Lower Bound for a Fixed Experiment Selection Strategy}\label{lbfixedexp}
In this sub-section, we will fix the experiment selection strategy to be $g$ and analyze the problem of optimizing the inference strategy for this particular experiment selection strategy. This analysis will help us in obtaining a lower bound on the misclassification probability and in designing inference strategies for Problem (\ref{opti}). Consider the following optimization problem.
\begin{align}\tag{P3}\label{opt3}
    & & & \underset{f \in \mathcal{F}}{\text{min}} & & \phi_N(i) & \\
    \nonumber & & & \text{subject to} & & \psi_N(i) \geq 1-\epsilon_N. & 
\end{align}
To analyze problem (\ref{opt3}),  we will first define some useful quantities related to the confidence level in Definition \ref{confidencelevel}.

%
%
%
%
%

For the hypothesis $i$ and a strategy $g \in \mathcal{G}$, define the likelihood distributions $P^g_{N,i}$ and $Q_{N,i}^g$ over the set $\mathcal{I}_{N+1}$ as follows
\begin{align}
P^g_{N,i}(\I_{N+1}) &\doteq \Py^g[I_{N+1} = \I_{N+1} \mid \rv{X} = i]\\
Q^g_{N,i}(\I_{N+1}) &\doteq \Py^g[I_{N+1}  = \I_{N+1}\mid \rv{X} \neq i].
\end{align}
\begin{proposition}\label{prop1}
Under any experiment selection strategy $g$, with probability 1, we have
\begin{align}
\log\frac{P^g_{N,i}(I_{N+1})}{Q^g_{N,i}(I_{N+1})} = \mathcal{C}_{i}(\vct{\rho}_{N+1})- \mathcal{C}_{i}(\vct{\rho}_1).\label{confllr}
\end{align}
\end{proposition}
\begin{IEEEproof}
See Appendix \ref{prop1proof}.
\end{IEEEproof}
Thus, the increment in confidence level is a log-likelihood ratio.
\begin{definition}[Expected Confidence Rate]\label{expconf}
We define the expected confidence rate $J_N^g(i)$ as
\begin{align}\label{eq:jng}
&J_N^g(i) \doteq \frac{1}{N} \E_i^{g} \left[\mathcal{C}_{i}(\vct{\rho}_{N+1})- \mathcal{C}_{i}(\vct{\rho}_1) \right].
\end{align}
\end{definition}
\begin{remark}
Due to Proposition \ref{prop1}, the expected confidence rate is the averaged Kullback-Leibler divergence between the distributions $P_{N,i}^g$ and $Q_{N,i}^g$. That is,
\begin{align*}
J_N^g(i) = \frac{1}{N}D(P_{N,i}^g || Q_{N,i}^g).
\end{align*}
\end{remark}

When the experiment selection strategy is fixed, we can view the problem (\ref{opti}) as a one-shot hypothesis testing problem in which we are trying to infer whether the collection of actions and observations $I_{N+1}$ is drawn from distribution $P^g_{N,i}$ or $Q^g_{N,i}$. This interpretation allows us to use the classical results \cite{cover2012elements,polyanotes} on one-shot hypothesis testing and derive various properties.

We will first obtain a lower bound on the misclassification probability $\phi_N(i)$ in Problem (\ref{opt3}) using the data-processing inequality of Kullback-Leibler divergences \cite{polyanotes}. This is commonly known as the weak converse \cite{liu2018second,polyanotes}.

\begin{lemma}[Weak Converse]\label{cslb}
Let $g$ be any given experiment selection strategy. Then for any inference strategy $f$ such that
$\psi_N(i) \geq  1-\epsilon_N,$ we have
\begin{align}\label{eq:cslb}
   -\frac{1}{N}\log\phi_N(i) &\leq \frac{J_N^g(i)}{1-\epsilon_N} + \frac{\log 2}{N(1-\epsilon_N)},
\end{align}
where $J_N^g(i)$ is the expected confidence rate. Therefore,
\begin{align}
-\frac{1}{N}\log\phi^*_N(i) 
 &\leq \frac{\sup_{g \in \mathcal{G}}J_N^g(i)}{1-\epsilon_N} + \frac{\log 2}{N(1-\epsilon_N)},\label{supbound}
\end{align}
where $\phi_N^*(i)$ is the optimum value in Problem (\ref{opti}).
\end{lemma}
\begin{IEEEproof}
See Appendix \ref{cslbproof}. Note that this lemma is true for every $0 \leq \epsilon_N < 1$. 
\end{IEEEproof}
The bound (\ref{supbound}) suggests that we can obtain a strategy-independent lower bound on $\phi_N^*$ by obtaining upper bounds on the quantity $\sup_{g \in \mathcal{G}}J_N^g(i)$. In the next sub-section, we will focus on obtaining this upper bound.

\subsection{Strategy-Independent Lower Bound}
We will first describe some important properties of the confidence level $\mathcal{C}_i(\vct{\rho})$ which will be used to derive a strategy-independent lower bound on the misclassifcation probability in problem (\ref{opti}). 

\begin{definition}\label{zdef}
For a given experiment selection strategy $g$ and an alternate hypothesis $j \neq i$, define the total log-likelihood ratio up to time $n$ as
\begin{align*}
\rv{Z}_n(j) \doteq  \sum_{k=1}^n\lambda_j^i(\rv{U}_k,\rv{Y}_k),
\end{align*}
where the log-likelihood ratio $\lambda_j^i$ is as defined in equation (\ref{llrdef}). Also, let
\begin{align*}
\bar{Z}_n \doteq  \sum_{j\neq i}\beta^{i*}(j)Z_n(j),
\end{align*}
\end{definition}
where $\beta^{i*}$ is the min-maximizing distribution in Definition \ref{kldef}.
Notice that the processes $Z_n(j)$ for each $j\neq i$ and $\bar{Z}_n$ are sub-martingales with respect to the filtration $I_{n+1}$ when $X =i$.
We will now establish the relationship between the total log-likelihood ratios $Z_n(j)$ and the confidence level $\mathcal{C}_i$.

\begin{lemma}\label{logsumexplemma}
For any experiment selection strategy $g$ and for each $1 \leq n \leq N$, we have
\begin{align}
\nonumber\mathcal{C}_i(\vct{\rho}_{n+1}) &- \mathcal{C}_i(\vct{\rho}_1) \\
&=  - \log\left[\sum_{j\neq i}\exp\Big(\log\tilde{\rho}_1(j) -  Z_n(j)\Big)\right],
\end{align}
where $\tilde{\rho}_1(j) = \rho_1(j)/(1-\rho_1(i)).$
\end{lemma}
\begin{IEEEproof}
This is a consequence of simple algebraic manipulations. See Appendix \ref{logsumexplemmaproof} for details.
\end{IEEEproof}

Note that for any vector $z$, we have $-\log\sum_j\exp(-z(j)) \approx \min_j z(j)$. Thus, the interpretation of this lemma is that the increment in confidence level $\mathcal{C}_i(\vct{\rho}_{N+1}) - \mathcal{C}_i(\vct{\rho}_1)$ approximately represents the smallest total log-likelihood ratio $\min_{j\neq i} Z_N(j)$. Therefore, this lemma can be seen as the first step towards establishing the relationship between the average expected increment in confidence $J_N^g$ and the max-min Kullback-Leibler divergence $D^*(i)$. To formally establish this relationship, we use Lemma  \ref{logsumexplemma} to decompose the increment in confidence into a non-positive cross-entropy term and a sub-martingale. This decomposition will be used in deriving strategy-independent lower bounds, both weak and strong, on the misclassification probability in Problem (\ref{opti}).

\begin{lemma}[Decomposition]\label{decom}
For any experiment selection strategy $g$, we have
\begin{align*}
&\mathcal{C}_i(\vct{\rho}_{n+1}) - \mathcal{C}_i(\vct{\rho}_1) = -H(\beta^{i*},\tilde{\rho}_{n+1}) + \bar{Z}_n + H(\beta^{i*},\tilde{\rho}_1).
\end{align*}
Here, $\tilde{\rho}_n(j) = \rho_n(j)/(1-\rho_n(i)).$ As a result of the non-negativity of cross entropy, we have
\begin{align}
\mathcal{C}_i(\vct{\rho}_{n+1}) - \mathcal{C}_i(\vct{\rho}_1) \leq \bar{Z}_n + H(\beta^{i*},\tilde{\rho}_1).
\end{align}
\end{lemma}
\begin{IEEEproof}
This is an algebraic consequence of Lemma \ref{logsumexplemma}. See Appendix \ref{decomproof} for details.
\end{IEEEproof}


Using Lemma \ref{decom}, we will now establish the relationship between the confidence rate $J_N^g(i)$ and the max-min Kullback-Leibler divergence $D^*(i)$ defined in equation (\ref{maxmin}). This, in conjunction with Lemma \ref{cslb}, will give us a strategy-independent lower bound on $\phi_N^*(i)$.
\begin{lemma}\label{kldbound}
For any experiment selection strategy $g$, we have
\begin{align}
    J_N^g(i) \leq D^*(i) + \frac{H(\beta^{i*},\tilde{\rho}_1)}{N},
\end{align}
where $\tilde{\rho}_1(j) = \rho_1(j)/(1-\rho_1(i)).$ Further, using Lemma \ref{cslb} and Assumption \ref{epsassum}, we can conclude that
\begin{align}
\label{limsuplb}\limsup_{N \to \infty}\;-\frac{1}{N}\log{\phi_N^*(i)} \leq D^*(i).
\end{align}
\end{lemma}

\begin{IEEEproof}
See Appendix \ref{klboundproof}.
\end{IEEEproof}

\subsection{Achievability of Decay-Rate $D^*(i)$ in Problem (\ref{opti})}\label{achievable}

We will now construct inference and experiment selection strategies that satisfy the constraint on hit probability $\psi_N(i)$ and asymptotically achieve misclassification decay-rate of $D^*(i)$. We will begin with the construction and analysis of the inference strategy.


The following is an upper bound on the misclassification probability associated with a deterministic confidence-threshold based inference strategy of the form discussed in Section \ref{mainresults}. Incidentally, this bound does not depend on the experiment selection strategy $g$.

\begin{lemma}\label{thresh}
Let $f$ be a deterministic inference strategy in which hypothesis $i$ is decided only if $\mathcal{C}_i(\vct{\rho}_{N+1}) - \mathcal{C}_i(\vct{\rho}_1) \geq \theta$. Then $\phi_N(i) \leq e^{-\theta}.$
\end{lemma}
\begin{IEEEproof}
See Appendix \ref{threshproof}.
\end{IEEEproof}
The inference strategy $f^N$ is constructed as follows
\begin{align}
{f}^N(\vct{\rho}_{N+1}:i) = 
\begin{cases}
1 &\text{if } \mathcal{C}_i(\vct{\rho}_{N+1}) - \mathcal{C}_i(\vct{\rho}_{1}) \geq  \theta_N\\
0 &\text{otherwise},
\end{cases}
\end{align}
where $\theta_N = ND^*(i) - o(N)$ and its precise value is provided in equation \eqref{thetadefsym} in Appendix \ref{detthmproof}. 
Due to Lemma \ref{thresh}, we can say that under the inference strategy $f^N$, any experiment selection strategy $g$ achieves achieves $\phi_N(i) \leq e^{-\theta_N}.$
However, the inference strategy and the experiment selection strategy must also satisfy the constraint $\psi_N(i) \geq 1 - \epsilon_N$. In Section \ref{thm2subsec}, we discussed experiment selection strategies that satisfy Criterion \ref{crit}. Let $g^N$ be any such experiment selection strategy that satisfies Criterion \ref{crit}. Let the type-$i$ correct-inference and misclassification probabilities associated with the strategy pair $(g^N,f^N)$ be $\psi_N(i)$ and $\phi_N(i)$, respectively. We can show that this strategy pair satisfies the constraint $\psi_N(i) \geq 1 - \epsilon_N$. The proof is in Appendix \ref{detthmproof}. 
Using the result \eqref{limsuplb}, Lemma \ref{thresh} and the fact that $\theta_N /N \to D^*(i)$ as $N \to \infty$, we can say that
\begin{align}
\lim_{N \to \infty}\frac{1}{N}\log\frac{1}{\phi_N(i)} = D^*(i).
\end{align} 
Thus, the experiment selection strategies that satisfy Criterion \ref{crit} (including ORS, DAS and DAS-RS discussed in Section \ref{thm2subsec}), when used in conjunction with the inference strategy $f^N$ described above, are feasible solutions for the optimization problem \eqref{opti} and asymptotically achieve a type-$i$ misclassification probability decay rate of $D^*(i)$. This concludes the proof of Theorem \ref{detthm}.

Since the inference strategy $f^N$ and the experiment selection strategy $g^N$ described above are feasible strategies with respect to the optimization problem in \eqref{opti}, we can say that $\phi_N^*(i) \leq \phi_N(i)  \leq e^{-\theta_N}.$
And since under Assumption \ref{epsassum}, $\theta_N/N \to D^*(i)$ as $N \to \infty$,
we have
\begin{align}
    \liminf_{N \to \infty}-\frac{1}{N}\log\phi^*_N(i) \geq  D^*(i).
\end{align}
Combining this result with the upper bound on the asymptotic decay rate of $\phi_N^*(i)$ in Lemma \ref{kldbound}, we have Theorem \ref{p2asym}.



\subsection{Tighter Non-asymptotic Lower Bounds}
In this section, we will provide an alternate approach to finding lower bounds on the misclassification probability $\phi_N(i)$. This approach can be used to obtain tight lower bounds in some special cases. We will later illustrate this procedure with the help of an example. 

For any given pair of inference and experiment selection strategies $f,g$ that are feasible in Problem (\ref{opti}), recall that the increment in confidence can be viewed as a log-likelihood ratio (\ref{confllr}). Therefore for this strategy pair $f,g$, we have the following for every $\chi \in \R$
\begin{align}
&-\log{\phi_N(i)}\\
 &\stackrel{a}{\leq}\, \chi - \log(\psi_N(i) - \Py_i^g[\mathcal{C}_i(\vct{\rho}_{N+1}) - \mathcal{C}_i(\vct{\rho}_1) > \chi])\\
& \stackrel{b}{\leq} \, \chi - \log(1-\epsilon_N - \Py_i^g[\mathcal{C}_i(\vct{\rho}_{N+1}) - \mathcal{C}_i(\vct{\rho}_1) > \chi])\\
&= \, \chi - \log(\Py_i^g[\mathcal{C}_i(\vct{\rho}_{N+1}) - \mathcal{C}_i(\vct{\rho}_1) \leq \chi]-\epsilon_N )\label{strong3}
\end{align}
Here, we use the convention that if $x \leq 0$, then $\log x \doteq -\infty$. Inequality $(a)$ is a consequence of the strong converse theorem in \cite{polyanotes}. Inequality $(b)$ holds because $\psi_N(i) \geq 1-\epsilon_N$. However, much like the weak converse in Lemma \ref{cslb}, this lower bound on $\phi_N(i)$ depends on the experiment selection strategy $g$. We made use of the decomposition in Lemma \ref{decom} to obtain a strategy-independent lower bound in Lemma \ref{kldbound}. We will follow a similar approach here.
We have
\begin{align}
&\Py^g_i[\mathcal{C}_i(\vct{\rho}_{N+1}) - \mathcal{C}_i(\vct{\rho}_1) \leq \chi] \\
&\stackrel{a}{=} \Py^g_i[-H(\beta^{i*},\tilde{\rho}_{N+1})+\bar{Z}_N + H(\beta^{i*},\tilde{\rho}_1) \leq \chi]\\
&\stackrel{b}{\geq} \Py^g_i[\bar{Z}_N + H(\beta^{i*},\tilde{\rho}_1) \leq \chi].\label{strong1}
\end{align}
Equality $(a)$ is a consequence of Lemma \ref{decom}, and since $H(\beta^{i*},\tilde{\rho}_{N+1}) \geq 0$, we have that the event
\begin{align}
\{-H(\beta^{i*},\tilde{\rho}_{N+1})+\bar{Z}_N + H(\beta^{i*},\tilde{\rho}_1) \leq \chi\} \\
\supseteq \{\bar{Z}_N + H(\beta^{i*},\tilde{\rho}_1) \leq \chi\},\label{strong2}
\end{align}
which results in the inequality $(b)$. Combining \eqref{strong3} and \eqref{strong1} leads us to the following lemma.
\begin{lemma}[Stong Converse]\label{strongconv}
For any given pair of inference and experiment selection strategies $f,g$ that are feasible in Problem (\ref{opti}), we have for every $\chi \in \R$
\begin{align*}
-\log{\phi_N(i)} \leq \chi - \log(\Py^g_i[\bar{Z}_N + H(\beta^{i*},\tilde{\rho}_1) \leq \chi]-\epsilon_N ),
\end{align*}
with the convention that $\log x \doteq -\infty$ if $x \leq 0$.
\end{lemma}

Note that this lower bound is also dependent on the strategy $g$. However, it may be easier to derive strategy-independent lower bounds using the bound in Lemma \ref{strongconv}. This is because the process $\bar{Z}_n - nD^*(i)$ is a super-martingale given $X = i$. In fact, if every experiment in $\U$ is in the support of $\alpha^{i*}$ (which is the case in many problems), then the process $\bar{Z}_n - nD^*(i)$ is a martingale given $X = i$. Thus, a lower bound on $\phi_N(i)$ may be obtained using a strategy-independent lower bound on the tail probability $\Py^g_i[\bar{Z}_N + H(\beta^{i*},\tilde{\rho}_1) \leq \chi]$ \cite{freedman1975tail}. 
In some special cases, it may even occur that the evolution of the process $\bar{Z}_n$ is completely independent of the strategy $g$. We will discuss an example that satisfies this condition in Section \ref{sec:example}.

\section{Analysis of Problem (\ref{opt1})}\label{thm3proof}
In this section, we will analyze Problem \eqref{opt1} and prove Theorem \ref{mainthm3}.
%
%
%
%
Let $f,g$ be inference and experiment selection strategies that satisfy the constraints in Problem \eqref{opt1}, \emph{i.e.} $\psi_N(i) \geq 1- \epsilon_N$ for every $i \in \mathcal{X}$, where $\psi_N(i)$ is the type-$i$ correct-inference probability associated with strategies $f,g$. Let $\phi_N(i)$ be the type-$i$ misclassification probability associated with strategies $f,g$. Since the strategy pair $f,g$ satisfies $\psi_N(i) \geq 1-\epsilon_N$, we can use Lemma \ref{cslb} to obtain the following inequality
\begin{align}
   -\frac{1}{N}\log\phi_N(i) &\leq \frac{J_N^g(i)+ \frac{\log 2}{N}}{1-\epsilon_N} ,
\end{align}
for every $i \in \mathcal{X}$. Let $\gamma_N$ be the misclassification probability associated with strategy pair $f,g$. Then, we have
\begin{align}
    \gamma_N  &= \sum_{i \in \mathcal{X}}(1-\rho_1(i))\phi_N(i)\\
    &\geq \sum_{i \in \mathcal{X}}(1-\rho_1(i))\exp\left(-\frac{(NJ_N^g(i)+ \log 2)}{1-\epsilon_N} \right).
\end{align}
Rearranging the terms above, we have
\begin{align}
&-\frac{1}{N}\log \gamma_N\\
 &{\leq} -\frac{1}{N}\log\sum_{i \in \mathcal{X}}(1-\rho_1(i))e^{\left(-\frac{(NJ_N^g(i) + \log 2)}{1-\epsilon_N} \right)}\\
&= -\frac{1}{N}\log\sum_{i \in \mathcal{X}}e^{\left(-\frac{(NJ_N^g(i) + \log 2)}{1-\epsilon_N}  + \log(1-\rho_1(i))\right)}\\
&\stackrel{a}{\leq} -\frac{1}{N}\max_{i \in \mathcal{X}}\left(-\frac{(NJ_N^g(i)+\log 2)}{1-\epsilon_N} + \log(1-\rho_1(i))\right)\\
&= \min_{i \in \mathcal{X}}\left(\frac{J_N^g(i)+ \frac{\log 2}{N}}{1-\epsilon_N} - \frac{\log(1-\rho_1(i))}{N}\right),\label{gammaineq}
\end{align}
where inequality $(a)$ is because $\log\sum_{i}\exp(x_i) \geq \max_i(x_i)$. Using the definition of $\gamma_N^*$, we have
\begin{align}
&-\frac{1}{N}\log \gamma_N^*\\
&=\sup_{f,g: \psi_N(i) \geq 1-\epsilon_N,i\in\mathcal{X}}\left[-\frac{1}{N}\log \gamma_N\right]\\
&\stackrel{b}{\leq}\min_{i \in \mathcal{X}} \left(\frac{\sup_{g}J_N^g(i)+ \frac{\log 2}{N}}{1-\epsilon_N} - \frac{\log(1-\rho_1(i))}{N}\right)\\
&\stackrel{c}{\leq} \min_{i \in \mathcal{X}}\left(\frac{D^*(i)+  \frac{H(\beta^{i*},\tilde{\rho}_1) + \log 2 }{N}}{1-\epsilon_N}  - \frac{\log(1-\rho_1(i))}{N}\right).
\end{align}
Inequality $(b)$ is due to the result in \eqref{gammaineq} and inequality $(c)$ follows from Lemma \ref{kldbound}. Thus, we can conclude that
\begin{align}\label{lbthm}
\limsup_{N \to \infty}-\frac{1}{N}\log \gamma_N^* \leq \min_{i \in \mathcal{X}}D^*(i).
\end{align}
This establishes an upper bound on the asymptotic decay rate of the optimal misclassification probability $\gamma_N^*$ in Problem \eqref{opt1}. We will now show that this rate is asymptotically achievable by constructing  appropriate experiment selection and inference strategies.


\subsubsection{Experiment selection strategy}\label{stratsubsubsec} For a given horizon $N$ and for each $i \in \mathcal{X}$, let $g^{N,i}$ be an experiment selction strategy that satisfies Criterion \ref{crit} with respect to hypothesis $i$. Let the \emph{maximum likelihood} (ML) estimate at time $n$ be \begin{equation}\bar{i}_n \doteq \argmax_{i \in \mathcal{X}}\bar{\rho}_n(i),\end{equation}where ties are broken arbitrarily in the $\argmax$ operator and $\bar{\rho}_n$ is the posterior belief at time $n$ formed using uniform prior at time 1 instead of $\rho_1$. If $\bar{i}_n = i$, then an experiment is selected randomly with distribution $g_n^{N,i}(I_n)$. We denote this experiment selection strategy by $\bar{g}^N$. Note that if $g^{N,i}$ is ORS, then the strategy $\bar{g}^N$ is identical to the one in \cite{chernoff1959sequential}.
\subsubsection{Inference strategy} Consider the following deterministic inference strategy $\bar{f}^N$ where for each $i \in \mathcal{X}$
$$
\bar{f}^N(\vct{\rho}_{N+1}:i) = 
\begin{cases}
1 &\text{if } \mathcal{C}_i(\vct{\rho}_{N+1}) - \mathcal{C}_i(\vct{\rho}_{1}) \geq \theta_N(i)\\
0 &\text{otherwise},
\end{cases}
$$
where $-\mathcal{C}_i(\rho_1) < \theta_N(i) = ND^*(i) - o(N)$ and is precisely defined in Appendix \ref{achievesymproof}. Notice that if $\bar{f}^N(\vct{\rho}_{N+1}:i) = 0$ for every $i \in \mathcal{X}$, then $\bar{f}^N(\vct{\rho}_{N+1}:\aleph) =1$.

Since $\theta_N(i) > 0$ for every $i \in \mathcal{X}$, the threshold condition in $\bar{f}^N$ can be satisfied by at most one hypothesis. Thus, $\bar{f}^N$ declares hypothesis $i$ if and only if the confidence increment associated with $i$ exceeds the threshold $\theta_N(i)$. Hence, for each hypothesis $i \in \mathcal{X}$, the inference strategy $\bar{f}^N$ admits the structure required for Lemma \ref{thresh} and thus, using Lemma \ref{thresh}, we can conclude that for each hypothesis $i \in \mathcal{X}$, $\phi_N(i) \leq e^{-\theta_N(i)}$. Therefore, under the strategies $(\bar{f}^N,\bar{g}^N)$, we have
\begin{align}
    \gamma_N \leq \sum_{i \in \mathcal{X}}(1-\rho_1(i))\exp(-\theta_N(i)).
\end{align}
In Appendix \ref{achievesymproof}, we show that there exists an integer $\bar{N}$ such that for every $N \geq \bar{N}$, the strategy pair $(\bar{f}^N, \bar{g}^N)$ also satisfies all the type-$i$ correct-inference probability constraints in problem (\ref{opt1}). Thus, for every $N \geq \bar{N}$, we have
$$\gamma_N^* \leq \gamma_N \leq \sum_{i \in \mathcal{X}}(1-\rho_1(i))\exp(-\theta_N(i)),$$
and hence,
\begin{align}
\label{achieve}&\liminf_{N \to \infty}-\frac{1}{N}\log\gamma_N^* \\
&\geq \lim_{N \to \infty} - \frac{1}{N}\log \sum_{i \in \mathcal{X}}(1-\rho_1(i))\exp(-\theta_N(i))\\
& \nonumber\geq \lim_{N \to \infty} -\frac{1}{N} \log \Big( M\max_{i \in \mathcal{X}}\{(1-\rho_1(i))\exp(-\theta_N(i))\} \Big)\\
&= \lim_{N \to \infty} -\frac{1}{N}  \Big( \max_{i \in \mathcal{X}}\{\log\big((1-\rho_1(i))\exp(-\theta_N(i))\big)\} \Big)\\
&= \min_{i \in \mathcal{X}} \left\{ \lim_{N\to \infty}\frac{\theta_N(i) - \log(1-\rho_1(i))}{N}\right\}\\
&= \min_{i\in \mathcal{X}}D^*(i).
\end{align}


Using the results \eqref{lbthm} and \eqref{achieve}, we can conclude that
\begin{align}
\lim_{N \to \infty}-\frac{1}{N}\log\gamma_N^* = \min_{i\in \mathcal{X}}D^*(i).
\end{align}
This concludes the proof of Theorem \ref{mainthm3}.

\section{An Example: Anomaly Detection}\label{sec:example}
Consider a system with two sensors $\mathscr{A}$ and $\mathscr{B}$. These sensors can detect an anomaly in the system in their proximity. The system state $X$ can take three values $\{0,1,2\}$ where $X = 0$ indicates that the system is safe, \emph{i.e.} there is no anomaly in the system. On the other hand, $X = 1$ indicates that there is an anomaly near sensor $\mathscr{A}$ and $X = 2$ indicates that there is an anomaly near sensor $\mathscr{B}$. The prior belief $\rho_1$ is uniform over the set $\{0,1,2\}$. There is a controller that can activate one of these sensors at each time to obtain an observation. Thus, the collection of actions that the controller can select from is $\U = \{\mathscr{A},\mathscr{B}\}$. The observations are binary, i.e. $\mathcal{Y} = \{0,1\}$. The conditional probabilities $\Py[Y = 1 \mid X,U]$ associated with the observations given various states and actions are given in Table \ref{table1}.

\begin{figure}
\centering
\begin{tabular}{ | l | c | c | c |}
 \hline
   & $X = 0$& $X= 1$ & $X = 2$ \\
  \hline
  $U = \mathscr{A}$ & $1-\nu$ & $\nu$ & $1-\nu$ \\
  $U = \mathscr{B}$ & $1-\nu$ & $1-\nu$ & $\nu$ \\
  \hline
\end{tabular}
\caption{Conditional probabilities $\Py[Y = 1 \mid X, U]$. In our numerical experiments $\nu = 0.6$ which indicates that the observations from these sensors are very noisy.}
\label{table1}
\end{figure}

\subsubsection{Formulation}\label{anomalyform}
After collecting $N$ observations from these sensors, we are interested in determining whether the system is safe or unsafe. We consider a setting where incorrectly declaring the system to be safe can be very expensive while a few false alarms can be tolerated. In this setting, the inconclusive decision $\aleph$ is treated as an alarm. Therefore, we would like to design an experiment (sensor) selection strategy $g$ and an inference strategy $f$ that minimize the probability $\phi_N(0)$ of incorrectly declaring the system to be safe subject to the condition that the probability $\psi_N(0)$ of correctly declaring the system to be safe is sufficiently high. This can be formulated as
\begin{align}\tag{P4}\label{opt4}
    & & & \underset{f \in \mathcal{F},g \in \mathcal{G}}{\text{min}} & & \phi_N(0) & \\
    \nonumber & & & \text{subject to} & & \psi_N(0) \geq 1-\epsilon_N, & 
\end{align}
where $\epsilon_N = \min\{0.05,10/N\}$ in our numerical experiments. Notice that this fits the formulation of Problem \eqref{opti}.

\subsubsection{Asymptotically Optimal Rate and Weak Converse}
Using Theorem \ref{p2asym}, we can conclude that the asymptotically optimal misclassification rate is $$D^*(0) = \max_{\vct{\alpha} \in \Delta\mathcal{U}} \min_{j\neq 0} \sum_{u}\alpha(u) D(p_0^u || p_j^u) = (\nu - 1/2)\log\frac{\nu}{1-\nu}.$$
The max-minimizer $\alpha^{0*}(\mathscr{A}) = \alpha^{0*}(\mathscr{B}) = 0.5$ and the min-maximizer $\beta^{0*}(1) = \beta^{0*}(2) = 0.5$. For convenience, we will refer to $\alpha^{0*}$ as $\alpha^*$ and $\beta^{0*}$ as $\beta^*$. Also, notice that the bound on the log-likelihood ratios $B = \log\frac{\nu}{1-\nu}$, $\nu > 0.5$. Therefore using the weak converse in Lemma \ref{cslb} and Lemma \ref{kldbound}, we have
\begin{align*}
\frac{1}{N}\log\frac{1}{\phi^*_N(0)} \leq  \left(\frac{\nu - 1/2}{1-\epsilon_N} \right)\log\frac{\nu}{1-\nu} + \frac{2\log2}{N(1-\epsilon_N)}.
\end{align*}

\subsubsection{Strong Converse}
We will now use the lower bound in Lemma \ref{strongconv} to derive a strategy-independent lower bound one $\phi_N(0)$. Define
\begin{align}
L_n \doteq \beta^*(1)\lambda^0_1(U_n,Y_n) + \beta^*(2)\lambda^0_2(U_n,Y_n)
\end{align}
Notice that given $X = 0$, for \emph{either} experiment $u \in \mathcal{U}$, we have
$$
\beta^*(1)\lambda^0_1(u,Y) + \beta^*(2)\lambda^0_2(u,Y) = 
\begin{cases}
\frac{1}{2}\log\frac{\nu}{1-\nu} & \text{w.p. } \nu\\
\frac{1}{2}\log\frac{1-\nu}{\nu} & \text{w.p. } 1-\nu.
\end{cases}
$$
Let the moment generating function of the variable above be $\bar{\mu}(s)$. Therefore, for any strategy $g$, we have
\begin{align}
&\E_0^g[\exp(\sum_{k =1}^ns_kL_k)] = \E_0^g[\E_0\exp(\sum_{k =1}^ns_kL_k)\mid I_n]]\\
&=\nonumber \E_0^g[\exp(\sum_{k =1}^{n-1}s_kL_k)\E_0[\exp(s_nL_n)\mid I_n]]\\
&= \E_0^g[\exp(\sum_{k =1}^{n-1}s_kL_k)]\bar{\mu}(s_n) =\Pi_{k=1}^n\bar{\mu}(s_k).
\end{align}
Thus, $L_n$ is an i.i.d. sequence and the process $\bar{Z}_n$ is the sum of these i.i.d. random variables. We can also exploit the fact that the observations are binary valued. Let the number of $0$'s  in $N$ observations be $K$. Then $K$ has binomial distribution with parameters $N,\nu$. Further, 
\begin{align}\label{vartobin}
\bar{Z}_N \leq \chi -\log 2 \iff \left(K-\frac{N}{2}\right)\log\frac{\nu}{1-\nu}  \leq \chi - \log 2.
\end{align}
Define
$$\chi^* = (\mathscr{Q}(2\epsilon_N) -N/2)\log\frac{\nu}{1-\nu} + \log 2,$$
where $\mathscr{Q}$ is the quantile function of the binomial distribution with parameters $N,\nu$. Using the relation (\ref{vartobin}), we can conclude that
\begin{align}
\Py^g_0[\bar{Z}_N + H(\beta^{i*},\tilde{\rho}_1) \leq \chi^*] \geq 2\epsilon_N,
\end{align}
under any experiment selection strategy $g$. Finally, using Lemma \ref{strongconv}, we have for any pair of inference and experiment selections strategies that satisfy the constraints in Problem \eqref{opt4}
\begin{align}
\log\frac{1}{\phi_N(0)} \leq \chi^* - \log(\epsilon_N).
\end{align}

\subsubsection{Deterministic Adaptive Strategy (DAS)}
If the sensor $\mathscr{A}$ is selected, then the log-likehood ratio $\lambda^0_2(\mathscr{A},Y)$ is identically 0. Therefore, $\mu^0_2(\mathscr{A},s) = 1$ for every $s$. Further, it can easily be verified that $\mu^0_1(\mathscr{A},s) = 1$ at $s = 0$ and $s=1$. This fact combined with the convexity of $\mu_1^0(\mathscr{A},s)$ implies that $\mu^0_1(\mathscr{A},s) \leq 1$ for every $0 \leq s \leq 1$. Similarly, $\mu^0_1(\mathscr{B},s) = 1$ for every $s$ and $\mu^0_2(\mathscr{B},s) \leq 1$ for every $0 \leq s \leq 1$. Because of this, the deterministic adaptive experiment selection strategy (DAS) in Section \ref{thm2subsec} reduces to the following:
\begin{align}
g_n(\rho_n: \mathscr{A}) =
\begin{cases}
1 & \text{if } \rho_n(1) \geq \rho_n(2)\\
0 & \text{otherwise}.
\end{cases}
\end{align}
Due to Theorem \ref{detthm}, we know this strategy is asymptoticaly optimal. Note that the strategy described above is independent of the time-horizon $N$.

\subsubsection{Numerical Results}
\begin{figure}
\centering
\includegraphics[width = \columnwidth]{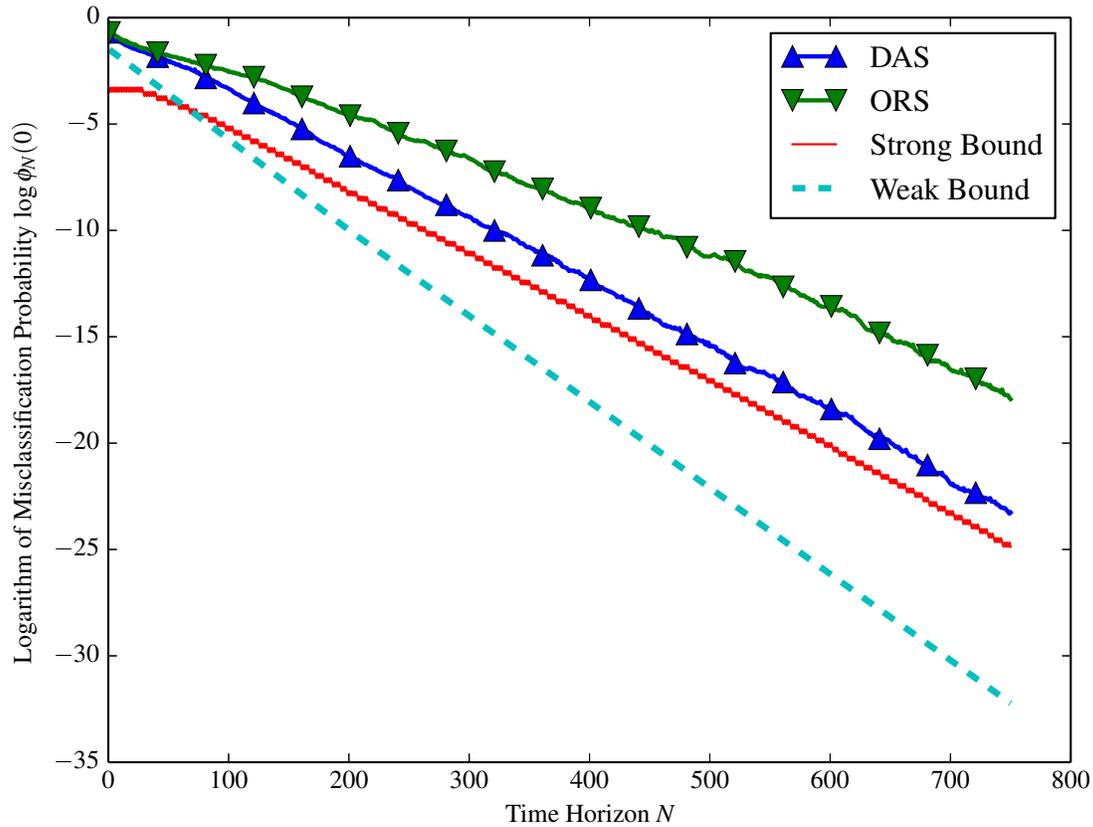}
\caption{The plot depicts the performance of strategies ORS and DAS. Both are asymptotically optimal but DAS is better in the non-asymptotic regime. When the horizon $N = 500$, we see a 13dB improvement in the misclassification probability with DAS. Also notice that the strong bound on misclassification probability is very close to the performance of DAS.}
\label{fig:plot1}
\end{figure}

We compare the performance of the open-loop randomized strategy $\alpha^{0*}$ and the deterministic adaptive strategy described above in Figure \ref{fig:plot1}. We observe that the performance of DAS is significantly better in the non-asymptotic regime. We also plot the weak and strong bounds established earlier. We observe that the strong bound is very close to the performance of the deterministic adaptive strategy.

In our numerical experiments, the inference strategy is a confidence threshold based strategy. Instead of computing the threshold, we empirically find the best threshold using a binary search.

\begin{remark}
Directly evaluating the misclassifcation probability $\phi_N(i)$ is computationally infeasible because $\phi_N(i)$ decreases exponentially and is of the order $10^{-5}$ to $10^{-10}$. To counter this issue we use the following method to approximately compute $\phi_N(i)$. We can show using basic algebra (see Lemma \ref{approxlemma} in Appendix \ref{auxres}) that
\begin{align*}
&\log\frac{1}{\phi_N(i)} \\
&\stackrel{a}{=} -\log\E^g_i[e^{-(\mathcal{C}_{i}(\vct{\rho}_{N+1})- \mathcal{C}_{i}(\vct{\rho}_1) - \log	 \mathbbm{1}_{\mathcal{A}_{N+1}}(I_{N+1}))}]\\
& \stackrel{b}{\approx} \log |\mathscr{S}| -\log\sum_{\mathscr{S}}[e^{-(\mathcal{C}_{i}(\vct{\rho}_{N+1})- \mathcal{C}_{i}(\vct{\rho}_1) - \log	 \mathbbm{1}_{\mathcal{A}_{N+1}}(I_{N+1}))}],\\
\end{align*}
where $\mathcal{A}_{N+1}$ is the region of acceptance of hypothesis $i$ and $\mathbbm{1}_{\mathcal{A}_{N+1}}(I_{N+1})$ takes the value 1 if ${I}_{N+1} \in \mathcal{A}_{N+1}$ and $0$ otherwise. The approximation $(b)$ is obtained by replacing the expectation in $(a)$ with the sample average. Note that $\mathscr{S}$ denotes the collection of sampled trajectories. The advantage of making this approximation is that the expression in $(b)$ involves a log-sum-exp function which is computationally stable and avoids underflows. This log-sum-exp function can be updated iteratively as well.
\end{remark}

\subsubsection{Chernoff's Deterministic Strategy}\label{chern}
In \cite{chernoff1959sequential}, Chernoff described a fully deterministic strategy for \emph{sequential} hypothesis testing but gave an example scenario for which his strategy was \emph{not} asymptotically optimal (see Section 7, \cite{chernoff1959sequential}). We will now demonstrate that even in such pathological scenarios, our strategies designed based on Criterion \ref{crit} are asymptotically optimal and also, have a better performance in the non-asymptotic regime.

Consider the same anomaly detection setup and formulation as in Section \ref{anomalyform} with two additional experiments $\mathscr{C}$ and $\mathscr{D}$. The conditional distributions of the observations associated with these experiments are provided in Table \ref{table2}. Chernoff's approach to deterministic strategy design is as follows. Let $\bar{j}_n$ be the most-likely \emph{alternate} hypothesis at time $n$, \emph{i.e.}
\begin{align}
\bar{j}_n \doteq \argmax_{j \neq 0} \bar{\rho}_n(j),
\end{align}
where $\bar{\rho}_n$ is the posterior belief at time $n$ formed using a uniform prior. Then at time $n$, perform the experiment that maximizes $D(p_0^u || p_{\bar{j}_n}^u)$. In other words, select the experiment that can best distinguish the most-likely alternate hypothesis from the hypothesis of interest (in this case, it is $X =0$). In our setup, Chernoff's deterministic strategy reduces to the following
\begin{align}
U_n =
\begin{cases}
\mathscr{A} & \text{if } \rho_n(1) \geq \rho_n(2)\\
\mathscr{B} & \text{otherwise}.
\end{cases}
\end{align}
It can be shown that this strategy is \emph{not} asymptotically optimal. Chernoff's randomized strategy (ORS) in this case reduces to selecting either experiment $\mathscr{C}$ or $\mathscr{D}$ with probability 0.5.

\begin{figure}
\centering
\includegraphics[width = \columnwidth]{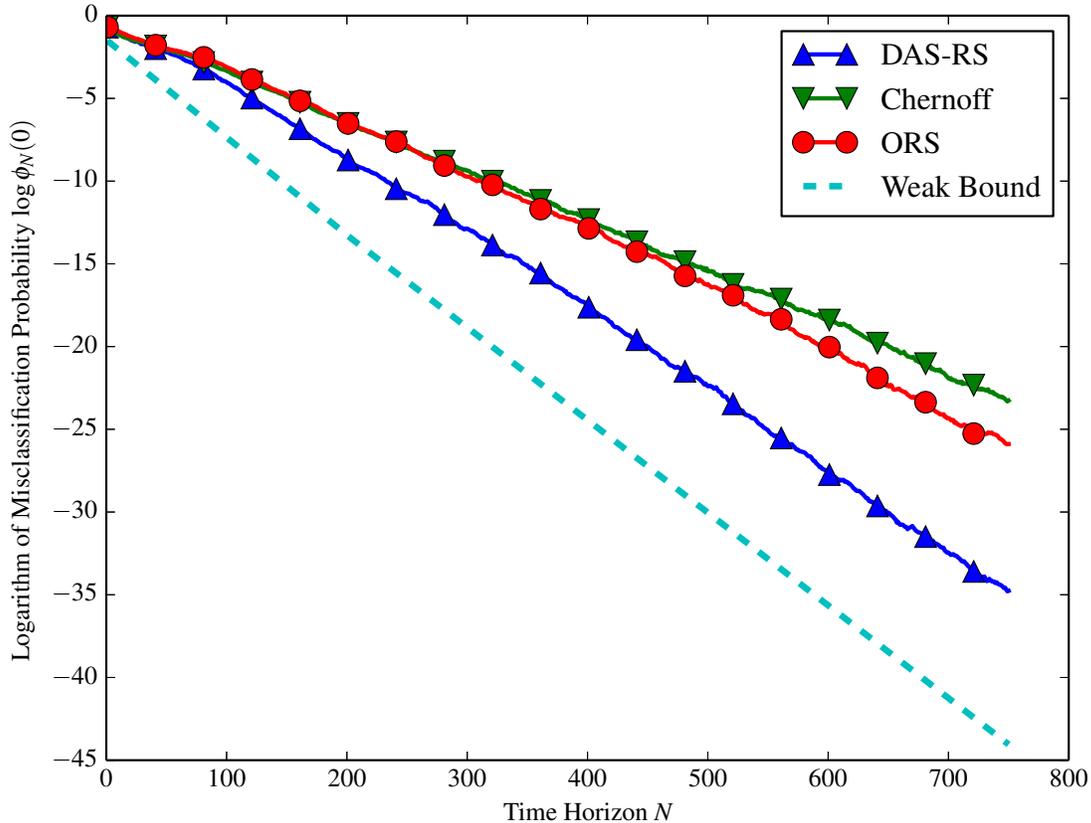}
\caption{Type-0 misclassification probability associated with strategies DAS-RS, ORS and Chernoff's deterministic strategy. Both ORS and DAS-RS are asymptotically optimal and Chernoff's deterministic strategy is not. Notice that DAS-RS outperforms all the other strategies in the non-asymptotic regime.}
\label{fig:plot2}
\end{figure}

On the other hand, with some simple calculations, we can show that our strategy DAS-RS described in Section \ref{thm2subsec} reduces to
\begin{align}
U_n =
\begin{cases}
\mathscr{C} & \text{if } \rho_n(1) \geq \rho_n(2)\\
\mathscr{D} & \text{otherwise}.
\end{cases}
\end{align}
Because of Theorem \ref{detthm}, we know that both ORS and DAS-RS are asymptotically optimal. The performances of ORS, DAS-RS and Chernoff's deterministic strategy are shown in Figure \ref{fig:plot2}. In this case, the strategy DAS-RS does not depend on the time-horizon $N$ whereas the strategy DAS depends on $N$.

\begin{figure}
\centering
\begin{tabular}{ | l | c | c | c |}
 \hline
   & $X = 0$& $X= 1$ & $X = 2$ \\
  \hline
  $U = \mathscr{A}$ & $0.400$ & $0.600$ & $0.400$ \\
  $U = \mathscr{B}$ & $0.400$ & $0.400$ & $0.600$ \\
  $U = \mathscr{C}$ & $0.402$ & $0.598$ & $0.280$ \\
  $U = \mathscr{D}$ & $0.402$ & $0.280$ & $0.598$ \\
  \hline
\end{tabular}
\caption{Conditional probabilities $\Py[Y = 1 \mid X, U]$ for the problem setup in Section \ref{chern}.}
\label{table2}
\end{figure}

\section{Conclusion}\label{conc}
We formulated two fixed horizon active hypothesis testing problems (asymmetric and symmetric) in which the agent can decide on one of the hypotheses or declare its experiments inconclusive. Using information-theoretic techniques, we obtained lower bounds on optimal misclassification probability in these problems. We also derived upper bounds by constructing appropriate strategies and analyzing their performance. We proposed a novel approach to designing deterministic and adaptive strategies for these active hypothesis testing problems. We proved that these deterministic strategies are asymptotically optimal, and through numerical experiments, demonstrated that they have significantly better performance in the non-asymptotic regime in some problems of interest.

\appendices
\section{Auxiliary Results}\label{auxres}
\begin{proposition}\label{beltolog}
Let $i \in \mathcal{X}$ be a hypothesis. For any $j \neq i$ and at each time $n$, let $\tilde{\rho}_{n}(j) = \rho_n(j)/ (1-\rho_n(i))$. Then for any experiment selection strategy $g$, we have
\begin{align}
\tilde{\rho}_{n}(j) = \frac{e^{(\log \tilde{\rho}_1(j) - Z_{n-1}(j)})}{\sum_{k \neq i}e^{(\log \tilde{\rho}_1(k) - Z_{n-1}(k)})},
\end{align}
with probability 1. Here, $Z_n(j)$ is the total log-likelihood ratio defined in Definition \ref{zdef}.
\end{proposition}
\begin{IEEEproof}
We have
\begin{align*}
&\tilde{\rho}_n(j) = \frac{\rho_n(j)}{\sum_{k \neq i}\rho_n(k)}=\frac{\rho_1(j)\prod_{m=1}^{n-1}p_j^{\rv{U}_m}(\rv{Y}_m)}{\sum_{k\neq i}\rho_1(k)\prod_{m=1}^{n-1}p_k^{\rv{U}_m}(\rv{Y}_m)}\\
&=\frac{\tilde{\rho}_1(j)\prod_{m=1}^{n-1}\frac{p_j^{\rv{U}_m}(\rv{Y}_m)}{p_i^{\rv{U}_m}(\rv{Y}_m)}}{\sum_{k\neq i}\tilde{\rho}_1(k)\prod_{m=1}^{n-1}\frac{p_k^{\rv{U}_m}(\rv{Y}_m)}{p_i^{\rv{U}_m}(\rv{Y}_m)}}=  \frac{e^{(\log \tilde{\rho}_1(j) - Z_{n-1}(j)})}{\sum_{k \neq i}e^{(\log \tilde{\rho}_1(k) - Z_{n-1}(k)})}.
\end{align*}
\end{IEEEproof}
\begin{corollary}\label{scor}
Under the setting of Proposition \ref{beltolog}, we have for each $0 \leq s \leq 1$ and $j \neq i$,
\begin{align}
\frac{(\rho_n(j))^s}{\sum_{k \neq i}(\rho_n(k))^s} =  \frac{e^{(s\log \tilde{\rho}_1(j) - sZ_{n-1}(j)})}{\sum_{k \neq i}e^{(s\log \tilde{\rho}_1(k) - sZ_{n-1}(k)})}.
\end{align}
\end{corollary}
\begin{IEEEproof}
This result is a consequence of Proposition \ref{beltolog} and some simple algebraic manipulations.
\end{IEEEproof}

\begin{lemma}\label{approxlemma}
Let $g$ be any experiment selection strategy and let $f$ be a deterministic inference strategy. Let $\mathcal{A}_{N+1}$ be the region in which the inference policy $f$ selects hypothesis $i$, that is
$$\mathcal{A}_{N+1} := \{\I: f(\I:i) = 1 , \Py^{f,g}[\rv{I}_{N+1} = \I] \neq 0\}. $$
Then
\begin{align*}
\phi_N(i) = \E_i^g\left[e^{(-(\mathcal{C}_{i}(\vct{\rho}_{N+1})- \mathcal{C}_{i}(\vct{\rho}_1) - \log	 \mathbbm{1}_{\mathcal{A}_{N+1}}(I_{N+1}))} \right].
\end{align*}
\end{lemma}
\begin{IEEEproof}
Using Proposition \ref{prop1}, we have
\begin{align*}
&\E_i^g\left[e^{(-(\mathcal{C}_{i}(\vct{\rho}_{N+1})- \mathcal{C}_{i}(\vct{\rho}_1) - \log	 \mathbbm{1}_{\mathcal{A}_{N+1}}(I_{N+1})))} \right]\\
&=\E_i^g\left[\exp{\left(-\log\frac{P^g_{N,i}(I_{N+1})}{Q^g_{N,i}(I_{N+1})}  + \log	 \mathbbm{1}_{\mathcal{A}_{N+1}}(I_{N+1})\right)} \right]\\
&=\sum_{\I \in \mathcal{A}_{N+1}}P^g_{N,i}(\I) \frac{Q^g_{N,i}(\I)}{P^g_{N,i}(\I)}  = \sum_{\I \in \mathcal{A}_{N+1}}Q^g_{N,i}(\I) = \phi_N(i).
\end{align*}
\end{IEEEproof}

\section{Proof of Proposition \ref{prop1}}\label{prop1proof}
For any instance $\I_{N+1} \in \mathcal{I}_{N+1}$ such that $\Py^g[I_{N+1} = \I_{N+1}] > 0$, we have the following using Bayes' rule
\begin{align}
\log\frac{P^g_{N,i}(\I_{N+1})}{Q^g_{N,i}(\I_{N+1})} &=  \log\frac{\Py^g[I_{N+1}= \I_{N+1} \mid \rv{X} = i]}{\Py^g[I_{N+1}= \I_{N+1} \mid X\neq i]}\label{rationonzero}\\
&= \mathcal{C}_{i}(\vct{\rho}_{N+1})- \mathcal{C}_{i}(\vct{\rho}_1),\label{confeqllr}
\end{align}
where $\rho_{N+1}$ is the posterior belief associated with the instance $\I_{N+1}$. Note that because of Assumption \ref{boundedassump}, $\Py^g[I_{N+1} = \I_{N+1}] > 0$ implies that both the numerator and the denominator in (\ref{rationonzero}) are non-zero and thus, the expression in (\ref{rationonzero}) is well-defined. Since equation (\ref{confeqllr}) is true for every instance $\I_{N+1}$ with non-zero probability, we have our result.

\section{Proof of Lemma \ref{cslb}}\label{cslbproof}
Define a random variable $X^\dagger$ as follows: $X^\dagger \doteq 1$ if $\hat{X}_{N} = i$ and $X^\dagger \doteq 0$ otherwise.
Thus, we have
\begin{align}
\psi_N(i) &=\Py^{f,g}[X^\dagger = 1 \mid X = i]\\
\phi_N(i) &= \Py^{f,g}[X^\dagger = 1 \mid X \neq i].
\end{align}
In this proof, let us denote $\psi_N(i)$ with $\psi$ and $\phi_N(i)$ with $\phi$ for convenience. Notice that under strategies $g$ and $f$, the variables $X, I_{N+1}$ and $X^\dagger$ form a Markov chain. That is,
\begin{align}
\Py^{f,g}[{X}^\dagger = 1 \mid X,I_{N+1}] = \Py^{f,g}[{X}^\dagger =1 \mid I_{N+1}].
\end{align}
Therefore, using the data-processing property of relative entropy \cite{polyanotes, liu2018second}, we can conclude that
\begin{align*}
D(P_{N,i}^g || Q_{N,i}^g) &\geq \psi \log\frac{\psi }{\phi} + (1-\psi )\log\frac{1-\psi }{1-\phi}\\
&\stackrel{a}{\geq} -\psi\log{\phi} +\psi\log{\psi} +(1-\psi)\log{(1-\psi)}\\
&\stackrel{b}{\geq} -\psi\log{\phi} - \log 2\\
& \stackrel{c}{\geq} -(1-\epsilon_N)\log{\phi} - \log 2.
\end{align*}
Inequality $(a)$ follows from the fact that $1 - \phi \leq 1$. Inequality $(b)$ holds because $-\psi\log{\psi} -(1-\psi)\log{(1-\psi)}$ is a binary entropy and can at most be $\log 2$. Inequality $(c)$ follows from our assertion that $\psi_N(i) \geq 1-\epsilon_N$. Therefore, we have
\begin{align}
-\frac{1}{N}\log\phi_N(i) &\leq \frac{J_N^g(i)}{1-\epsilon_N} + \frac{\log 2}{N(1-\epsilon_N)}.
\end{align}

\section{Proof of Lemma \ref{logsumexplemma}}\label{logsumexplemmaproof}
 We have
 \begin{align}
 &\log\frac{\rho_{n+1}(i)}{1-\rho_{n+1}(i)}-\log\frac{\rho_1(i)}{1-\rho_1(i)} \\
 \stackrel{a}{=} &\log \frac{\rho_1(i)\prod_{m=1}^np_i^{\rv{U}_m}(\rv{Y}_m)}{\sum_{j\neq i}\rho_1(j)\prod_{m=1}^np_j^{\rv{U}_m}(\rv{Y}_m)}-\log\frac{\rho_1(i)}{1-\rho_1(i)} \\
 = &\log \frac{\prod_{m=1}^np_i^{\rv{U}_m}(\rv{Y}_m)}{\sum_{j\neq i}\tilde{\rho}_1(j)\prod_{m=1}^np_j^{\rv{U}_m}(\rv{Y}_m)} \\
 \nonumber = & - \log\sum_{j\neq i}\exp\Big(\log\tilde{\rho}_1(j) +  \sum_{m=1}^n\lambda_i^j(\rv{U}_m,\rv{Y}_m)\Big)\\
 &= - \log\left[\sum_{j\neq i}\exp\Big(\log\tilde{\rho}_1(j) -  Z_n(j)\Big)\right].
 \end{align}
 Equality $(a)$ follows from the fact that the observation $Y_m$ is independent of the past $I_m = \{U_{1:m-1},Y_{1:m-1} \}$ conditioned on the hypothesis $X$ and the current experiment $U_m$.

 \section{Proof of Lemma \ref{decom}}\label{decomproof}
Using the definition of cross-entropy, we have
\begin{align*}
&-H(\beta^{i*},\tilde{\rho}_{n+1}) = \sum_{j \neq i}\beta^{i*}(j)\log\tilde{\rho}_{n+1}(j)\\
&\stackrel{a}{=}\sum_{j \neq i}\beta^{i*}(j)\log\left(\frac{\tilde{\rho}_1(j)e^{-Z_n(j)}}{\sum_{k\neq i}\tilde{\rho}_1(k)e^{-Z_n(k)}} \right)\\
&= -H(\beta^{i*},\tilde{\rho}_1) - \bar{Z}_n - \log\left[\sum_{k\neq i}\exp\left(\log\tilde{\rho}_1(k) -  Z_n(k)\right)\right]\\
&\stackrel{b}{=}-H(\beta^{i*},\tilde{\rho}_1) - \bar{Z}_n + \mathcal{C}_i(\vct{\rho}_{n+1}) - \mathcal{C}_i(\vct{\rho}_1).
\end{align*}
Equality $(a)$ follows from Proposition \ref{beltolog} in Appendix \ref{auxres} and equality $(b)$ is a consequence of Lemma \ref{logsumexplemma}.

\section{Proof of Lemma \ref{kldbound}}\label{klboundproof}
Using the definition of expected confidence rate, we have
 \begin{align}
     \nonumber J_N^g(i)  &= \frac{1}{N} \E_i^{g} \left[\mathcal{C}_{i}(\vct{\rho}_{N+1})- \mathcal{C}_{i}(\vct{\rho}_1)\right]\\
    \nonumber   &\stackrel{a}{\leq} \frac{H(\beta^{i*},\tilde{\rho}_1)}{N} + \frac{1}{N}\E_i^g\left[\sum_{j\neq i}\beta^{i*}(j)\sum_{n=1}^N\lambda_j^i(\rv{U}_n,\rv{Y}_n)\right]\\
   \nonumber    &\stackrel{b}{=}  \frac{H(\beta^{i*},\tilde{\rho}_1)}{N} + \frac{1}{N}\E_i^g\left[\sum_{j\neq i}\beta^{i*}(j)\sum_{n=1}^ND(p_i^{\rv{U}_n}||p_j^{\rv{U}_n})\right]\\
 \nonumber      &=\frac{H(\beta^{i*},\tilde{\rho}_1)}{N} + \frac{1}{N}\E^g\left[\sum_{n=1}^N\sum_{j\neq i}\beta^{i*}(j)D(p_i^{\rv{U}_n}||p_j^{\rv{U}_n})\right]\\
  \nonumber     &\stackrel{c}{\leq} \frac{H(\beta^{i*},\tilde{\rho}_1)}{N} + \frac{1}{N}\E_i^g\left[\sum_{n=1}^N D^*(i)\right]\\
    \label{dstareq} &= \frac{H(\beta^{i*},\tilde{\rho}_1)}{N} + D^*(i).
 \end{align}
Equality $(a)$ is a consequence of Lemma \ref{decom}. Equality ($b$) follows from the fact that
  \begin{align}
     \E^g_i  \sum_{n=1}^N\lambda_j^i(\rv{U}_n,\rv{Y}_n) &= \E^g_i\sum_{n=1}^N\E_i[\lambda_j^i(\rv{U}_n,\rv{Y}_n)\mid \rv{U}_n]\\
     &= \E^g_i\sum_{n=1}^N D(p_i^{\rv{U}_n}||p_j^{\rv{U}_n}).
 \end{align}
 Inequality $(c)$ follows from the definition of the min-max distribution $\beta^{i*}$. Combining inequalities \eqref{dstareq} and \eqref{supbound} from Lemma \ref{cslb} gives us \eqref{limsuplb}.

\section{Proof of Lemma \ref{thresh}}\label{threshproof}
Let $\mathcal{A}_{N+1}$ be the region in which the inference policy $f$ described in Lemma \ref{thresh} selects hypothesis $i$, that is
$$\mathcal{A}_{N+1} := \{\I: f(\I:i) = 1 , \Py^{f,g}[\rv{I}_{N+1} = \I] \neq 0\}. $$
We have
\begin{align}
   \nonumber &\Py^{f,g}[\hat{\rv{X}}_{N+1} = i , \rv{X} \neq i ] = \Py^{g}[\rv{I}_{N+1} \in \mathcal{A}_{N+1} , \rv{X} \neq i ]\\
 \nonumber   &= \sum_{\I \in \mathcal{A}_{N+1}}\Py^{g}[\rv{I}_{N+1} = \I , \rv{X} \neq i ]\\
  \nonumber  &= \sum_{\I \in \mathcal{A}_{N+1}}\Py^{g}[\rv{I}_{N+1} = \I , \rv{X} = i ]e^{\left[-\log\frac{\Py^{g}[\rv{I}_{N+1} = \I , \rv{X} = i ]}{\Py^{g}[\rv{I}_{N+1} = \I , \rv{X} \neq i ]}\right]}\\
    &\stackrel{a}{=} \sum_{\I \in \mathcal{A}_{N+1}}\Py^{g}[\rv{I}_{N+1} = \I , \rv{X} = i ]\exp\left[-\mathcal{C}_i(\rho)\right]\label{confapproxeq}\\
    &\stackrel{b}{\leq} \sum_{\I \in \mathcal{A}_{N+1}}\Py^{g}[\rv{I}_{N+1} = \I , \rv{X} = i ]\exp\left[-(\theta + \mathcal{C}_i(\vct{\rho}_1))\right]\\
    &\stackrel{c}{\leq} \rho_1(i)e^{-(\theta + \mathcal{C}_i(\vct{\rho}_1))} = (1-\rho_1(i))e^{-\theta}.
\end{align}
In equality $(a)$, $\rho$ is the posterior belief on $X$ given information $\I$. Equality $(a)$ follows from the definition of confidence level. Inequality $(b)$ follows from the fact that $\mathcal{C}_i(\rho) \geq \theta + \mathcal{C}_i(\vct{\rho}_1)$ for every $\I \in \mathcal{A}_{N+1}$. And inequality $(c)$ is simply because $\Py^{g}[\rv{I}_{N+1} = \I , \rv{X} = i ] \leq \Py[ \rv{X} = i ]$. Therefore,
\begin{equation}
    \phi_N(i) = \Py^{f,g}[\hat{\rv{X}}_{N+1} = i \mid \rv{X} \neq i ] \leq e^{-\theta}.
\end{equation}

\section{Proof of Theorem \ref{detthm}}\label{detthmproof}

Let us fix the horizon $N$. In this proof, we will drop the superscript from $g^N$ and simply refer to it as $g$ for convenience. Since the inference strategy has a threshold structure (see \eqref{fNdef}), proving that $\psi_N(i) \geq 1 - \epsilon_N$ is equivalent to proving that the probability 
\begin{align*}
\Py_i^g[\mathcal{C}_i&(\vct{\rho}_{N+1}) - \mathcal{C}_i(\vct{\rho}_1) < \theta_N] \leq \epsilon_N.
\end{align*}
To this end, we will begin with obtaining upper bounds on the moment-generating function (MGF) of the confidence increment and then obtain a Chernoff bound based on these upper bounds. 

\subsubsection{Confidence Level and Log-likelihood Ratios}
Let $0 < s \leq 1$. Using Lemma \ref{logsumexplemma}, we have
\begin{align}
\nonumber\E^g_i\exp[&-s(\mathcal{C}_i(\vct{\rho}_{N+1}) - \mathcal{C}_i(\vct{\rho}_1))]\\
& = \E^g_i\left(\sum_{j\neq i}\exp\left(\log\tilde{\rho}_1(j) -  Z_N(j)\right)\right)^s\\
\nonumber &= \E^g_i\left(\sum_{j\neq i}(\exp\left(s\log\tilde{\rho}_1(j) -  sZ_N(j)\right))^{1/s}\right)^s\\
&\stackrel{a}{\leq}\E^g_i\left[\sum_{j\neq i}\exp\left(s\log\tilde{\rho}_1(j) -  sZ_N(j)\right)\right]\\
&= \sum_{j\neq i}\E^g_i\exp\left(s\log\tilde{\rho}_1(j) -  sZ_N(j)\right).\label{normineq}
\end{align}
Inequality $(a)$ holds because $\| \cdot \|_{1/s} \leq \| \cdot \|_1.$ Inequality \eqref{normineq} provides an upper bound on the MGF of the confidence increment in terms of the MGFs of the log-likelihood ratios.

\subsubsection{Open-loop Randomized Experiment Selection at time ${n+1}$}
Let $n < N$. Consider a scenario in which all the experiments upto time $n$ are selected using strategy $g$ but the experiment $U_{n+1}$ is randomly selected using the distribution $\alpha^{i*}$ instead of using the strategy $g$. Under this modified strategy (say $\tilde{g}$), for some alternate hypothesis $j \neq i$, we have
\begin{align}
&\E^{\tilde{g}}_i\exp\left(s\log\tilde{\rho}_1(j) -  sZ_{n+1}(j)\right)\\
&=\nonumber\E^g_i[\E^{\tilde{g}}_i[\exp\left(s\log\tilde{\rho}_1(j) -  sZ_{n+1}(j)\right) \mid I_{n+1}]]\\
&= \E^g_i[\exp\left(s\log\tilde{\rho}_1(j) -  sZ_{n}(j)\right)\label{condexpsres}\\
&\nonumber\qquad\qquad\qquad\qquad\times\E^{\tilde{g}}_i[\exp(-s\lambda_j^i(U_{n+1},Y_{n+1})) \mid I_{n+1}]].
\end{align}

Let us analyze the term $\E^{\tilde{g}}_i[\exp(-s\lambda_j^i(U_{n+1},Y_{n+1})) \mid I_{n+1}]$ in \eqref{mgfineq} separately. Since the observation $Y_{n+1}$ is conditionally independent of $I_{n+1}$ given the experiment $U_{n+1}$ and the true hypothesis $X$ (see equation (\ref{obs})), we have
\begin{align}
&\E_i^{\tilde{g}}[\exp(-s\lambda_j^i(U_{n+1},Y_{n+1})) \mid I_{n+1}] \label{mgf1}\\
\nonumber&= \sum_u \alpha^{i*}(u)\E_i[\exp(-s\lambda_j^i(u,Y))]= \sum_u \alpha^{i*}(u)\mu_j^i(u,s).
\end{align}
Also, under the same strategy $\alpha^{i*}$, the random variable $\lambda_j^i(U_{n+1},Y_{n+1})$ has mean $D^i_j \doteq \sum_{u}\alpha^{i*}D(p_i^u || p_j^u)$ and is bounded by $B$ with probability 1 because of Assumption \ref{boundedassump}. Bounded variables are sub-Gaussian (see Hoeffding's lemma in \cite{massart2007concentration}) and thus,  we have
\begin{align}
\nonumber\E_i^\alpha[\exp(-s\lambda_j^i(U_{n+1},&Y_{n+1})) \mid I_{n+1}] \\
&\leq \exp(-sD^i_j + s^2B^2/2)\\
&\leq \exp(-sD^*(i) + s^2B^2/2),\label{mgf2}
\end{align}
where the last inequality follows from the fact that $D^i_j \geq D^*(i)$ for all $j \neq i$. Combining the results (\ref{mgf1}) and (\ref{mgf2}), we can say that when $U_{n+1}$ is selected using $\alpha^{i*}$,
\begin{align}
\nonumber &\sum_{j\neq i}\exp\left(s\log\tilde{\rho}_1(j) -  sZ_{n}(j)\right)\\
& \qquad\qquad\times\E_i^{\tilde{g}}[\exp(-s\lambda_j^i(U_{n+1},Y_{n+1})) \mid I_{n+1}]\\
& = \sum_{j\neq i}\exp\left(s\log\tilde{\rho}_1(j) -  sZ_{n}(j)\right) \sum_u \alpha^{i*}(u)\mu_j^i(u,s)\\
& \leq \sum_{j\neq i}\exp\left(s\log\tilde{\rho}_1(j) -  sZ_{n}(j)\right) \exp(-sD^*(i) + s^2B^2/2).\label{mgfineq}
\end{align}
We will use the result in \eqref{mgfineq} to inductively obtain a bound on the MGF of the confidence increment under the strategy $g$.

\subsubsection{Inductive Step}
Having established the results \eqref{condexpsres} and \eqref{mgfineq}, we will now prove the following key Lemma.
\begin{lemma}\label{stratcondlemma}
For any experiment selection strategy $g$ that satisfies Criterion \ref{crit}, we have 
\begin{align}
\sum_{j \neq i}\E^g_i&\exp\Big(s\log\tilde{\rho}_1(j) -  sZ_{n+1}(j)\Big)\\
& \leq \left(\sum_{j \neq i}\E^g_i\exp\left(s\log\tilde{\rho}_1(j) -  sZ_{n}(j)\right)\right)\\
&\qquad \qquad \times \exp(-sD^*(i) + s^2B^2/2).
\end{align}
\end{lemma}
\begin{IEEEproof}
Since $g$ satisfies Criterion \ref{crit}, using Corollary \ref{scor} in Appendix \ref{auxres}, we have
\begin{align}
\nonumber&\frac{\sum_{j\neq i} \sum_u ({{\rho}_{n+1}(j)})^s\alpha_{n+1}(u)\mu_j^i(u,s)}{\sum_{j\neq i}  ({{\rho}_{n+1}(j)})^s} \\
\nonumber&\leq \frac{\sum_{j\neq i} \sum_u ({{\rho}_{n+1}(j)})^s\alpha^{i*}(u)\mu_j^i(u,s)}{\sum_{j\neq i}  ({{\rho}_{n+1}(j)})^s}\\
\nonumber\iff\; & \sum_{j\neq i}\exp\left(s\log\tilde{\rho}_1(j) -  sZ_{n}(j)\right) \sum_u \alpha_{n+1}(u)\mu_j^i(u,s)\\
&\leq \sum_{j\neq i}\exp\left(s\log\tilde{\rho}_1(j) -  sZ_{n}(j)\right) \sum_u \alpha^{i*}(u)\mu_j^i(u,s).\label{gsuff}
\end{align}
Recall that $\alpha_{n+1} \in \Delta \mathcal{U}$ is the distribution selected by the strategy $g$ at time $n+1$. Thus, we have
\begin{align}
&\sum_{j \neq i}\E^g_i\exp\left(s\log\tilde{\rho}_1(j) -  sZ_{n+1}(j)\right)\\
&=\sum_{j \neq i}\E^g_i[\E^g_i[\exp\left(s\log\tilde{\rho}_1(j) -  sZ_{n+1}(j)\right) \mid I_{n+1}]]\\
&= \sum_{j \neq i}\E^g_i[\exp\left(s\log\tilde{\rho}_1(j) -  sZ_{n}(j)\right)\\
&\nonumber\qquad\qquad\qquad\qquad\times\E^g_i[\exp(-s\lambda_j^i(U_{n+1},Y_{n+1})) \mid I_{n+1}]]\\
\nonumber&\stackrel{a}{=}  \E^g_i\sum_{j\neq i}\exp\left(s\log\tilde{\rho}_1(j) -  sZ_{n}(j)\right) \sum_u \alpha_{n+1}(u)\mu_j^i(u,s)\\
\nonumber&\stackrel{b}{\leq} \E^g_i\sum_{j\neq i}\exp\left(s\log\tilde{\rho}_1(j) -  sZ_{n}(j)\right) \sum_u \alpha^{i*}(u)\mu_j^i(u,s)\\
\nonumber&\stackrel{c}{\leq}\E^g_i\sum_{j\neq i}\exp\left(s\log\tilde{\rho}_1(j) -  sZ_{n}(j)\right) \\
&\qquad\qquad\qquad\qquad \times \exp(-sD^*(i) + s^2B^2/2),
\end{align}
{where equality $(a)$ follows from the fact that the observation $Y_{n+1}$ is conditionally independent of $I_{n+1}$ given $\alpha_{n+1}$.} Inequality $(b)$ follows from the result in \eqref{gsuff} and inequality $(c)$ is a consequence of the result \eqref{mgfineq}.
\end{IEEEproof}

Using the result in \eqref{normineq} and applying Lemma \ref{stratcondlemma} inductively, we have
\begin{align}
\E^g_i\exp[&-s(\mathcal{C}_i(\vct{\rho}_{N+1}) - \mathcal{C}_i(\vct{\rho}_1))]\\
& \leq \sum_{j\neq i}(\tilde{\rho}_1(j))^s\exp(-sND^*(i) + s^2NB^2/2)\\
&\leq M\exp(-sND^*(i) + s^2NB^2/2).\label{mgf3}
\end{align}

\subsubsection{Chernoff Bound}
We can use the Chernoff bound \cite{ross2014introduction} to conclude that
\begin{align}
\Py_i^g[\mathcal{C}_i&(\vct{\rho}_{N+1}) - \mathcal{C}_i(\vct{\rho}_1) < \theta]\\
& \leq \E^g_i\exp[-s(\mathcal{C}_i(\vct{\rho}_{N+1}) - \mathcal{C}_i(\vct{\rho}_1) -\theta)]\\
&\stackrel{a}{\leq} M\exp(s\theta-sND^*(i) + s^2NB^2/2)\\
&\stackrel{b}{=} \epsilon_N.
\end{align}
Inequality $(a)$ follows from the result in (\ref{mgf3}). Equality $(b)$ is obtained by substituting $\theta = \theta_N$ and $s = s_N$ where
\begin{align*}
\theta_N &\doteq ND^*(i) - \frac{s_N NB^2}{2}- {\frac{1}{s_N}\log\frac{M}{\epsilon_N}},\label{thetadef}
\end{align*}
and $s_N$ is as defined in \eqref{sdef}. Under Assumptions \ref{epsassum}, one can easily verify that $\theta_N/ N \to D^*(i)$ as $N \to \infty$.
Thus, we have shown that for the strategy pair $(f^N,g^N)$, $\psi_N(i) \geq 1-\epsilon_N$.


\section{Feasibility of Strategy in Section \ref{stratsubsubsec} for Problem \eqref{opt1}}\label{achievesymproof}
Let $\rv{T}$ be the smallest time index such that ML hypothesis $\bar{i}_n = X$ for every $n \geq \rv{T}$. That is,
\begin{align}
T = \min\{n': \bar{i}_n = X ~ \forall n \geq n'\}.
\end{align}
Notice that $\rv{T}$ is a random variable. Under Assumption \ref{steadinf}, it was shown in \cite{chernoff1959sequential} (Lemma 1) that there exist constants $b,K > 0$ such that for every $i\in \mathcal{X}$ and any strategy $g \in \mathcal{G}$, we have $\Py_i^g[\rv{T} > n] \leq Ke^{-bn}$. Let
\begin{align}
N' \doteq \left\lceil -\frac{1}{b}\log\frac{\epsilon_N}{2K} \right\rceil.
\end{align}
This ensures that  $\Py_i^g[\rv{T} > N'] \leq \epsilon_N/2$.
Fix a hypothesis $i$. Define the following event for each $n \geq N'$
\begin{align}
\mathscr{Z}_n = \{\bar{i}_k = X, N' \leq k \leq n \}.
\end{align}
Clearly, the events $\mathscr{Z}_n$ are decreasing with $n$. Also, we have
\begin{align}
\{ T \leq N'\} \subseteq \mathscr{Z}_n,
\end{align}
for every $n \geq N'$. 

Due to the threshold structure of the inference strategy $\bar{f}^N$, proving that $\psi_N(i) \geq 1 - \epsilon_N$ is equivalent to showing that 
\begin{align*}
\Py_i^g[\mathcal{C}_i&(\vct{\rho}_{N+1}) - \mathcal{C}_i(\vct{\rho}_1) < \theta_N(i)] \leq \epsilon_N.
\end{align*} To do so, we will use a Chernoff-bound based approach similar to the approach in Appendix \ref{detthmproof}. We have
\begin{align}
&\Py_i^g[\mathcal{C}_{i}(\vct{\rho}_{N+1})- \mathcal{C}_{i}(\vct{\rho}_1) < \theta_N(i)]\\
&= \Py_i^g[\mathcal{C}_{i}(\vct{\rho}_{N+1})- \mathcal{C}_{i}(\vct{\rho}_1) < \theta_N(i) , T > N']\\
&\quad + \Py_i^g[\mathcal{C}_{i}(\vct{\rho}_{N+1})- \mathcal{C}_{i}(\vct{\rho}_1) < \theta_N(i), T \leq N']\\
& \leq \epsilon_N/2 + \Py_i^g[\mathcal{C}_{i}(\vct{\rho}_{N+1})- \mathcal{C}_{i}(\vct{\rho}_1) < \theta_N(i), T \leq N'],\label{cont1}
\end{align}
where the last inequality follows from the definition of $N'$.

\subsubsection{Bounds on the MGF of Confidence Increment}
For some $0 \leq s \leq 1$, consider the following
\begin{align}
& \E^{\bar{g}}_i\exp[-s(\mathcal{C}_i(\vct{\rho}_{N+1}) - \mathcal{C}_i(\vct{\rho}_1)); T \leq N']\\
&\stackrel{a}{\leq}  \sum_{j\neq i}\E^{\bar{g}}_i[\exp\left(s\log\tilde{\rho}_1(j) -  sZ_N(j)\right); T \leq N']\\
& \stackrel{b}{\leq}  \sum_{j\neq i}\E^{\bar{g}}_i[\exp\left(s\log\tilde{\rho}_1(j) -  sZ_N(j)\right); \mathscr{Z}_N]\\
& \stackrel{c}{\leq}  \sum_{j\neq i}\E^{\bar{g}}_i[\exp\left(s\log\tilde{\rho}_1(j) -  sZ_{N-1}(j)\right); \mathscr{Z}_N]\\
&\qquad\qquad\qquad\times \exp(-sD^*(i) + s^2B^2/2)\\
& \stackrel{d}{\leq}  \sum_{j\neq i}\E^{\bar{g}}_i[\exp\left(s\log\tilde{\rho}_1(j) -  sZ_{N-1}(j)\right); \mathscr{Z}_{N-1}]\\
&\qquad\qquad\qquad\times \exp(-sD^*(i) + s^2B^2/2)\\
& \stackrel{e}{\leq}  \sum_{j\neq i}\E^{\bar{g}}_i[\exp\left(s\log\tilde{\rho}_1(j) -  sZ_{N'-1}(j)\right)]\\
&\qquad\qquad\qquad\times (\exp(-sD^*(i) + s^2B^2/2))^{N-N' +1}\\
& \stackrel{f}{\leq} M (\exp(-sD^*(i) + s^2B^2/2))^{N-N' +1}.\label{mgfsym1}
\end{align}
Inequality $(a)$ is a consequence of the result in equation \eqref{normineq}. Inequality $(b)$ holds because the event $\{ T \leq N' \} \subseteq \mathscr{Z}_N$. Notice that under the events $\mathscr{Z}_N$ and $\{X = i\}$, we have $\bar{i}_N = i$, and by the construction of strategy $\bar{g}^N$, the experiment $U_{N}$ is selected using the control law $g^{N,i}_N.$ This control law at time $N$ satisfies Criterion \ref{crit} which is the condition required for using Lemma \ref{stratcondlemma}. Thus, inequality $(c)$ can be obtained from the same arguments used to prove Lemma \ref{stratcondlemma}. Inequality $(d)$ holds because $\mathscr{Z}_{N} \subseteq \mathscr{Z}_{N-1}$. Inequality $(e)$ is obtained by inductively applying the arguments $(b) - (d)$. Inequality $(f)$ is due to Lemma \ref{lessthan1} state below.
\begin{lemma}\label{lessthan1}
Let $i, j \in \mathcal{X}$. If $0 \leq s \leq 1$, then under any strategy $g$ and for every $n$, we have
\begin{align}
\E^g_i[\exp\left(s\log\tilde{\rho}_1(j) -  sZ_{n}(j)\right)] \leq 1.
\end{align}
\end{lemma}
\begin{IEEEproof}
We have
\begin{align}
&\E^g_i[\exp\left(s\log\tilde{\rho}_1(j) -  sZ_{n}(j)\right)]\\
&= \E_i^g[\E^g_i[\exp\left(s\log\tilde{\rho}_1(j) -  sZ_{n}(j)\right)\mid I_n ]]\\
&= \E^g_i[\exp\left(s\log\tilde{\rho}_1(j) -  sZ_{n-1}(j)\right)]\mu_j^i(U_n,s)\\
&\stackrel{a}{\leq} \E^g_i[\exp\left(s\log\tilde{\rho}_1(j) -  sZ_{n-1}(j)\right)]\\
&\stackrel{b}{\leq} \E^g_i[\exp\left(s\log\tilde{\rho}_1(j) \right)] \leq 1.
\end{align}
Inequality $(a)$ is because for any experiment $u$, $\mu_j^i(u,s)$ is convex and $\mu_j^i(u,0) = \mu_j^i(u,1) = 1$. Inequality $(b)$ is obtained by inductively applying the same arguments.
\end{IEEEproof}

\subsubsection{Chernoff Bound}
Let $N'' \doteq N - N'+1$ and let $\zeta > 0$ be any small constant. Define
\begin{align}
\label{thetadefsym}&\theta_N(i) \\
\nonumber&\doteq \max \left\{\zeta-\mathcal{C}_i(\rho_1), N''D^*(i) - \frac{s_N N''B^2}{2}- {\frac{1}{s_N}\log\frac{2M}{\epsilon_N}}\right\},
\end{align}
where $s_N$ is as defined in \eqref{sdef}. Under Assumption \ref{epsassum}, one can verify that $\theta_N(i)/N \to D^*(i)$. Further, we can say that there exists an integer $\bar{N}$ such that $N'' > 0$ and for every $i \in \mathcal{X}$ and $N \geq \bar{N}$, we have $\theta_N(i) > \zeta -\mathcal{C}_i(\rho_1)$.
Thus, for every $N \geq \bar{N}$, using the Chernoff bound \cite{ross2014introduction}, we have
\begin{align*}
&\Py_i^g[\mathcal{C}_i(\vct{\rho}_{N+1}) - \mathcal{C}_i(\vct{\rho}_1) < \theta_N(i), T \leq N']\\
& \leq \E^g_i\exp[-s(\mathcal{C}_i(\vct{\rho}_{N+1}) - \mathcal{C}_i(\vct{\rho}_1) -\theta_N(i)); T \leq N']\\
&\stackrel{a}{\leq} M (\exp(s\theta_N(i)-sN''D^*(i) + s^2N''B^2/2)) \stackrel{b}{\leq} \epsilon_N/2,
\end{align*}
where inequality $(a)$ follows from \eqref{mgfsym1} and inequality $(b)$ is obtained by substituting the values of $\theta_N(i)$ and $s_N$. Combining this result with \eqref{cont1}, we have 
\begin{align*}
\Py_i^g[\mathcal{C}_i(\vct{\rho}_{N+1}) - \mathcal{C}_i(\vct{\rho}_1) \leq \theta_N(i)] \leq \epsilon_N.
\end{align*}

Therefore, the strategy pair $(\bar{f}^N,\bar{g}^N)$ defined in Section \ref{stratsubsubsec} satisfies the constraints in Problem \eqref{opt1}.

\section*{Acknowledgment}
This research was supported, in part, by National Science Foundation under Grant NSF CNS-1213128, CCF-1410009, CPS-1446901, Grant ONR N00014-15-1-2550, and Grant AFOSR FA9550-12-1-0215.

\ifCLASSOPTIONcaptionsoff
  \newpage
\fi



%
\bibliographystyle{IEEEtran}
\bibliography{refs}

%


\begin{IEEEbiographynophoto}{Dhruva Kartik} received the B.Tech. degree in electronics and communication engineering from the Indian Institute of Technology, Guwahati, India, in 2015. He joined the Ming Hsieh department of electrical and computer engineering at University of Southern California, Los Angeles, CA, USA in 2015 where he is working towards his Ph.D. degree. His research interests are in decentralized stochastic control, decision-making in sensing and communication systems, game theory, and reinforcement learning.

\end{IEEEbiographynophoto}


\begin{IEEEbiographynophoto}{Ashutosh Nayyar}
(S’09–M’11–SM’18) received the B.Tech.degree in electrical engineering from the Indian Institute of Technology, Delhi, India, in 2006. He received the M.S. degree in electrical engineering and computer science in 2008, the MS degree in applied mathematics in 2011, and the Ph.D. degree in electrical engineering and computer science in 2011, all from the University of Michigan, Ann Arbor. He was a Post-Doctoral Researcher at the University of Illinois at Urbana-Champaign and  at the University of California, Berkeley before joining the University of  Southern California in 2014. His research interests are in decentralized stochastic control, decentralized decision-making in sensing and  communication systems, game theory, mechanism design and electric energy systems. 
\end{IEEEbiographynophoto}

\begin{IEEEbiographynophoto}{Urbashi Mitra}
(F’07) received the B.S. and the M.S. degrees from the University of California at Berkeley and her Ph.D. from Princeton University.  Dr. Mitra is currently the Gordon S. Marshall Professor in Engineering at the University of Southern California. She was the inaugural Editor-in-Chief for the IEEE Transactions on Molecular, Biological and Multi-scale Communications. She has been a member of the IEEE Communication Society's Board of Governors (2018-2020), the IEEE Information Theory Society's Board of Governors (2002-2007, 2012-2017), the IEEE Signal Processing Society’s Technical Committee on Signal Processing for Communications and Networks (2012-2016), the IEEE Signal Processing Society’s Awards Board (2017-2018), and the Chair/Vice Chair of the IEEE Communications Society, Communication Theory Techcnial Committee(2019-2020,2017-2018). Dr. Mitra is a Fellow of the IEEE.  She is the recipient of: the 2017 IEEE Women in Communications Engineering Technical Achievement Award, a 2015 UK Royal Academy of Engineering Distinguished Visiting Professorship, a 2015 US Fulbright Scholar Award, a 2015-2016 UK Leverhulme Trust Visiting Professorship, IEEE Communications Society Distinguished Lecturer, 2012 Globecom Signal Processing for Communications Symposium Best Paper Award, 2012 US National Academy of Engineering Lillian Gilbreth Lectureship, the 2009 DCOSS Applications \& Systems Best Paper Award, 2001 Okawa Foundation Award, 2000 Ohio State University’s College of Engineering Lumley Award for Research, 1997 Ohio State University’s College of Engineering MacQuigg Award for Teaching, and a 1996 National Science Foundation CAREER Award.  She has been an Associate Editor for multiple IEEE publications. Dr. Mitra has held visiting appointments at: King’s College, London, Imperial College, the Delft University of Technology, Stanford University, Rice University, and the Eurecom Institute. Her research interests are in: wireless communications, communication and sensor networks, biological communication systems, detection and estimation and the interface of communication, sensing and control.
\end{IEEEbiographynophoto}




\end{document}